\documentclass[twocolumn]{aastex62}
\usepackage[]{threeparttable}
\usepackage{float}
\usepackage[caption = false]{subfig}
\usepackage{natbib}
\usepackage{appendix}
\usepackage{multirow}
\makeatletter
\newcommand\notsotiny{\@setfontsize\notsotiny{6}{7}}
\makeatother

\begin{document}
\title{The Elusive Distance Gradient in the Ultra-Faint Dwarf Galaxy Hercules: A Combined {\it Hubble Space Telescope} and {\it Gaia} View}

\correspondingauthor{Bur\c{c}in Mutlu-Pakdil}
\email{bmutlupakdil@as.arizona.edu}

\author[0000-0001-9649-4815]{BUR\c{C}\.{I}N MUTLU-PAKD\.{I}L}
\affil{Department of Astronomy/Steward Observatory, 933 North Cherry Avenue, 
Rm. N204, Tucson, AZ 85721-0065, USA}

\author[0000-0003-4102-380X]{DAVID J. SAND}
\affil{Department of Astronomy/Steward Observatory, 933 North Cherry Avenue, 
Rm. N204, Tucson, AZ 85721-0065, USA}

\author[0000-0002-1763-4128]{DENIJA CRNOJEVI\'{C}}
\affil{University of Tampa, 401 West Kennedy Boulevard, Tampa, FL 33606, USA}

\author[0000-0002-7157-500X]{EDWARD W. OLSZEWSKI}
\affil{Department of Astronomy/Steward Observatory, 933 North Cherry Avenue, Rm. N204, Tucson, AZ 85721-0065, USA}

\author[0000-0002-5177-727X]{DENNIS ZARITSKY}
\affil{Department of Astronomy/Steward Observatory, 933 North Cherry Avenue, Rm. N204, Tucson, AZ 85721-0065, USA}

\author[0000-0002-1468-9668]{JAY STRADER}
\affil{Department of Physics and Astronomy, Michigan State University,East Lansing, MI 48824, USA}

\author[0000-0002-1693-3265]{MICHELLE L. COLLINS}
\affil{Department of Physics, University of Surrey, Guildford GU2 7XH, UK}

\author[0000-0003-0248-5470]{ANIL C. SETH}
\affil{University of Utah, 115 South 1400 East Salt Lake City, UT 84112-0830, USA}

\author{BETH WILLMAN}
\affil{LSST and Steward Observatory, 933 North Cherry Avenue, Tucson, AZ 85721, USA}

\begin{abstract}

The ultra-faint dwarf galaxy Hercules has an extremely elongated morphology with both photometric overdensities and kinematic members at large radii, suggesting that it may be tidally disrupting due to a previous close encounter with the Milky Way. To explain its observational peculiarities, we present a deep {\it Hubble Space Telescope (HST)} imaging study of Hercules and its surrounding regions and investigate its tidal history through a careful search for a distance gradient along its stretched body. 
Our off-center {\it HST} data clearly resolve a main sequence, showing that the stellar extension seen along the major-axis of Hercules is genuine, not a clump of background galaxies. Utilizing {\it Gaia}~DR2 data, we clean the region around Hercules of field contamination, and find four new plausible member stars, all of which are located at the outskirts of the dwarf galaxy. We update the distance to Hercules, and find $130.6\pm1.2$~kpc ($m-M=20.58\pm0.02$) for the main body, which is consistent with earlier estimates in the literature. While we find no conclusive evidence for a distance gradient, our work demonstrates that constraining a distance gradient in such a faint system is not trivial, and the possible thickness of the dwarf along the line of sight and field contamination make it harder to make decisive conclusions even with these high-precision data. Future studies coupled with tailored theoretical models are needed to understand the true nature of Hercules and of tidal distortion observables in ultra-faint galaxies in general.

\end{abstract}


\section{Introduction}\label{sec:intro}

Ultra-faint dwarf galaxies (UFDs, $M_{V}\gtrsim-7$) represent the extreme end of the distribution of galaxy properties: the least luminous, least chemically enriched, and most dark matter dominated galaxies known. These systems offer a unique avenue to study the cosmological nature of dark matter and galaxy formation on the smallest scales \citep[for a recent review, see][]{BullockBoylan2017,Simon2019}. 

In recent years, deep wide-area photometric surveys have greatly increased the known population of Milky Way satellites \citep[e.g.,][]{Belokurov2008,Belokurov2009,Belokurov2010,Bechtol2015,Koposov2015,Koposov2018,DrlicaWagner2015,DrlicaWagner2016,Martin2015,KimJerjen2015,Kim2015,Laevens2015a,Laevens2015b,Torrealba2016,Homma2016,Homma2018,Mau2020}. Following their discovery, the first important step is to constrain their dynamical masses and dark matter content via their stellar velocity dispersions, but this heavily relies on the assumptions of dynamical equilibrium. Yet, both spectroscopic and deeper photometric follow up studies have uncovered signs of tidal interaction in several new ultra-faint systems \citep[e.g.,][among others]{Martin2008,Okamoto2008,Munoz2010,SG2007,Sand2009,Sand2012,Kirby2013,Munoz2018,Shipp2018,Li2018,Erkal2018,Carlin2018,Fritz2018,Longeard2018,MutluPakdil2018,Fu2019}. Therefore, in order to interpret them properly in a substructure formation context, it is crucial to determine whether UFDs are strongly affected by tidal forces. We recently addressed this fundamental question by presenting a comprehensive investigation of these signs of disruption in UFDs with a combined {\it HST}, MMT/Hectochelle, and {\it Gaia} study of the distant Milky Way UFD Leo~V \citep{MutluPakdil2019}. Our findings removed most of the observational clues that suggested Leo~V was disrupting, highlighting the importance of deeper studies into the nature of UFDs. In this paper, we extend our investigation to Hercules -- another strong candidate for a tidally disrupting UFD around the Milky Way -- with a different strategy, where we search for a distance gradient along the stretched body of the galaxy.

There is observational evidence that Hercules might be undergoing tidal disruption by the Milky Way. It has an extreme ellipticity ($\epsilon$$\sim0.7$, \citealt{Coleman2007,Martin2008,Sand2009}), making it the most elongated Milky Way satellite other than the disrupting Sagittarius dwarf \citep{Ibata1994}. Several photometric studies have found significant stellar overdensities far from its center \citep{Sand2009,Roderick2015}, and kinematics of a subset of member stars show velocity gradients \citep{Aden2009,Deason2012}. Additionally, \citet{Deason2012} found likely blue horizontal branch (HB) members at large distances, and \citet{Garling2018} identified RR Lyrae members outside the nominal tidal radius. 

Given the large distance of Hercules -- one of the outermost known Milky Way satellites at $D$$\approx$130~kpc -- it can only experience tidal stripping if its orbit is extremely eccentric, bringing it within $10-20$~kpc of the Galactic center. Such an orbit is plausible based on \textit{Gaia} proper motions, which variously\footnote{\citeauthor{Fritz2018} assumed a Milky Way dark matter halo with virial mass $1.6\times10^{12}M_{\odot}$ (or $0.8\times10^{12}M_{\odot}$). Both \citeauthor{Fu2019} and \citeauthor{Gregory2020} used a Milky Way virial mass $1.3\times10^{12}M_{\odot}$, but the latter also included the effect of the Large Magellanic Cloud and additional data points from new spectroscopy.} predict a pericenter of $14^{+23}_{-9}$~kpc (or $20^{+32}_{-14}$~kpc), \citep{Fritz2018}; $47^{+27}_{-21.6}$~kpc, \citep{Fu2019}; or $50.9^{+24.2}_{-23.6}$~kpc, \citep{Gregory2020}. Note that Hercules seems to be rapidly moving away from the Milky Way at $v_{GSR}\sim145$ km~s$^{-1}$ \citep{SG2007}, but its orbital path is not well constrained with existing proper motion data \citep{Gregory2020}.  

Two compelling orbital models have been put forward to explain the observed peculiarities of Hercules. One is the \citealt{MartinJin2010} hypothesis in which Hercules is a segment of a tidal stream observed near apocenter, and its elongation is aligned with its orbital path. This model provides two strong testable predictions: there should be a substantial distance and velocity gradient along the major-axis of Hercules. The other model is the \citealt{Kupper2017} `exploding satellite' scenario where the disruption of the satellite is caused by a close pericenter passage $\sim0.5$~Gyr ago. In this scenario, the stream-in-formation is actually aligned with the minor-axis of Hercules. In this case, there should be tidal debris along the dwarf's minor-axis and a distinct kinematic substructure, but the galaxy should not display any distance or velocity gradient along its extent. The positions of the substructures identified in \citet{Roderick2015} and distribution of RR Lyrae stars detected in \citet{Garling2018} are suggestive of an orbital path similar to this `exploding satellite' model. 

Combining available kinematic data with proper motions from {\it Gaia}~DR2, \citet{Fu2019} evaluated the probability that its orbit approaches sufficiently close to the Milky Way to experience tidal stripping, and found a probability of $\sim40$\% that Hercules has suffered tidal stripping. On the other hand, the authors were unable to confirm any members located in one of the most significant overdensities surrounding the galaxy that \citet{Roderick2015} identified ($\gtrsim3r_h$ at the West side of Hercules, see Figure~2 of \citealt{Fu2019}).  
Recently, \citet{Gregory2020} used new spectroscopy from DEIMOS/KeckII, together with the {\it Gaia}~DR2 data, and found no evidence for a significant velocity gradient or  velocity substructure in their membership sample. They also noted, however, that their average velocity uncertainty per member star is comparable to the overall Hercules velocity dispersion, and therefore may blur out any residual velocity substructures.
Additionally, the authors updated the systemic proper motion of Hercules, and found that the observed proper motion is slightly misaligned with the elongation of Hercules, in contrast to models which suggest that any tidal debris should be well aligned with the orbital path \citep[e.g.,][]{JinMartin2009,MartinJin2010}, but also inconsistent with the proper motion required for the `exploding satellite’ scenario \citep{Kupper2017}. However, they argued that the misalignment is not very significant, future observations from {\it Gaia}~DR3, the Rubin Observatory and the Roman Space Telescope may resolve this tension by providing significantly stronger constraints on the proper motion of Hercules. Fortunately, proper motion is not the only way to test the orbital models, and measurement of a distance gradient (or a lack thereof) across the body of Hercules may, in fact, be used to distinguish between these scenarios, and can serve as one of the most powerful probes for understanding the true nature of this object. 

In this work, we present a comprehensive deep {\it HST} imaging study of Hercules and its surrounding regions, combined with the {\it Gaia}~DR2 archival data, and search for observational evidence of tidal disturbance in the form of a distance gradient. We describe our observations and data reduction in Section~\ref{sec:obs}. We show its color-magnitude diagram (CMD) reaching well below the main sequence turn-off and compare it with that of M92 in Section~\ref{sec:cmd}. Using the {\it Gaia}~DR2 data, we perform a membership analysis where we search for new Hercules member candidates while reducing the foreground contamination in Section~\ref{sec:gaia}. In Section~\ref{sec:distance}, we revisit the distance to Hercules and explore the presence of a  distance gradient across the galaxy. Finally, we summarize our key results in Section~\ref{sec:conclusion}. 

\section{\textit{HST} Imaging and Data Reduction}\label{sec:obs}

\begin{figure}
\centering
\includegraphics[width=\columnwidth]{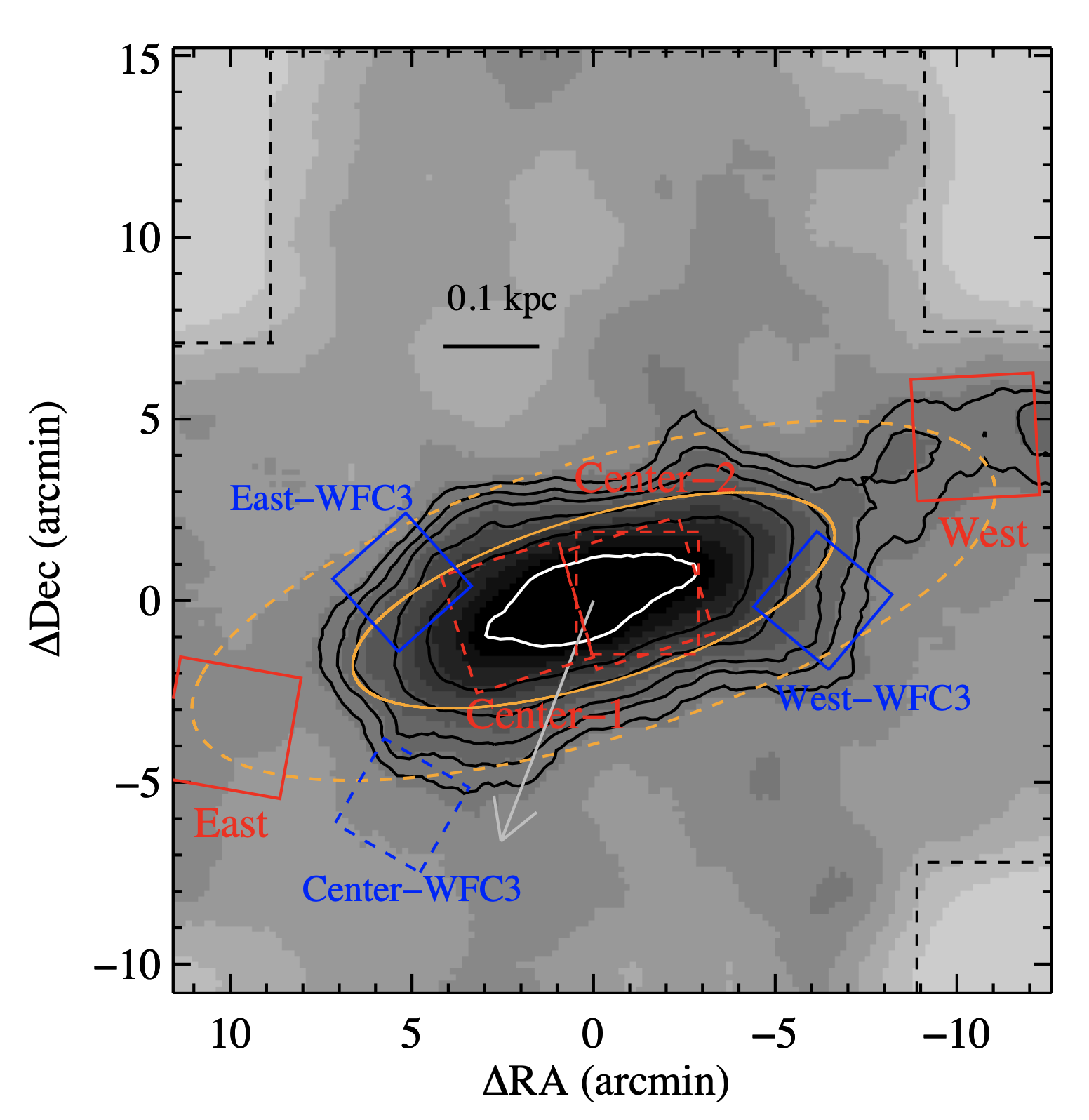}
\caption{Our observing strategy: the smoothed matched-filter map of Hercules is derived from the ground-based Large Binocular Telescope (LBT) imaging of \citet{Sand2009}, where the solid orange line marks the half-light radius (this value is comparable to the smoothing size). The dashed boxes represent the archived \textit{HST}/ACS and WFC3 imaging which we utilized for our analysis. The solid boxes are our \textit{HST}/ACS (red) observations along the major-axis and the relative WFC3 (blue) parallel ones. The orange dashed ellipse is the approximate isodensity contour at the radius of our ACS data ($r\sim2\times$half-light radii). The grey arrow marks the weighted mean proper motion of Hercules \citep{Gregory2020}. The black dashed lines represent the actual LBT field of view.
\label{fig:figure1}}
\end{figure}

Deep optical observations along the major-axis of Hercules were performed using the F606W and F814W filters on the Advanced Camera for Surveys (ACS) (HST-GO-15182; PI: D. Sand). 
Our observational strategy is outlined in Figure~\ref{fig:figure1}: the dashed boxes represent archived \textit{HST}/ACS and Wide Field Camera 3 (WFC3) imaging which we included in our analysis, and the solid boxes are our new observations. The $\sim$23~arcmin lever arm between our ACS pointings and the coordinated parallel observations with WFC3/UVIS were specifically designed to look for a predicted distance gradient across the galaxy. Table~\ref{tab:obslog} presents the log of the observations. A standard 4-point dither pattern was used to achieve 0.5 pixel sampling. The image depth was chosen to be consistent with the two available archival central pointings of Hercules from HST-GO-12549 (PI: T. Brown). 

We performed point-spread function photometry on all of the new and archival {\it HST} data as described in \citet{MutluPakdil2019}, which we briefly describe here. All photometry was performed on the flat-fielded (FLT) images using the latest version (2.0) of DOLPHOT \citep{Dolphin2002}, an updated version of HSTPHOT \citep{Dolphin2000}, largely using the recommended prescriptions on each camera. We used the synthetic Tiny Tim PSFs for all images. The catalogs were cleaned of background galaxies and stars with poor photometry, and we only included sources with (sharpness$_{F606W}+$sharpness$_{F814W})^2 < 0.1$, (crowding$_{F606W}+$crowding$_{F814W}) < 0.08$, signal-to-noise ratio $> 5$, roundness $< 1.5$, and object-type $\leq 2$ in each filter. We corrected for Milky Way extinction on a star-by-star basis using the \citet{Schlegel1998} reddening maps with the coefficients from \citet{Schlafly2011}. Tables~\ref{tab:C1}--\ref{tab:EWFC3} present our final catalogs, which include magnitudes (uncorrected for extinction) along with their DOLPHOT uncertainty, as well as the Galactic extinction values derived for each star. The extinction-corrected photometry is used throughout this work, and the CMDs are displayed in Figure~\ref{fig:figure2}. We derived completeness and photometric uncertainties using $\sim$50,000 artificial star tests per pointing (adding one
artificial star at a time), with the same routines used to create the photometric catalogs. 
Note that there are $\sim$6,000 stars in both our most populated central fields. 
The 50\% completeness limit for F814W is $\sim$28~mag across all {\it HST} images (see Table~\ref{tab:obslog}). 

\begin{table}
\centering
\small
\caption{Observation Log and Field Completeness of Hercules.} \label{tab:obslog}
\begin{tabular}{lccccc}
\tablewidth{0pt}
\hline
\hline
Field Name    & Camera &  Filter & Exp  & 50\%  & 90\%  \\
{}            &  {}    &  {}     & (s)  & (mag) & (mag) \\
\hline
Center-1 & ACS   &  F606W & 12880  & 28.55 & 27.96  \\
{} & ACS   &  F814W & 12745  & 28.12 & 27.58 \\
Center-2 & ACS   &  F606W & 12880  & 28.47 & 27.75  \\
{}  & ACS   &  F814W & 12745  & 28.05 & 27.27  \\
{Center}        & WFC3  &  F606W & 13635  & 28.17 & 27.15  \\
{}             & WFC3  &  F814W & 13515  & 27.97 & 27.36  \\
\hline
West & ACS   &  F606W & 12726  & 28.15 & 27.14 \\
{}    & ACS   &  F814W & 12726  & 27.90 & 27.34 \\
{}            & WFC3  &  F606W & 12926  & 28.15 & 27.02   \\
{}            & WFC3  &  F814W & 12926  & 27.91 & 27.36  \\
\hline
East & ACS   &  F606W & 13753  & 28.15 & 27.05   \\
{}    & ACS   &  F814W & 21763$^{\star}$  & 27.99 & 27.08 \\
{}            & WFC3  &  F606W & 12726  & 28.25 & 27.20  \\
{}            & WFC3  &  F814W & 20848$^{\star}$  & 28.02 & 27.37  \\
\hline
\end{tabular}
  \begin{tablenotes}
      \small
      \item Notes: The Center-1/ACS, Center-2/ACS and Center-WFC3 data are the available archival central fields of Hercules from HST-GO-12549 (PI: T. Brown). The rest is our new {\it HST} observations from HST-GO-15182. 
      \item $\star$ Due to gyro issues, we had to repeat the observations of this field, which provided some extra data. 
    \end{tablenotes}
\end{table}

\begin{table*}[tbp]
\caption{Photometry of Hercules Center~1-ACS.} \label{tab:C1}
\begin{minipage}[b]{0.95\linewidth}\centering
\begin{tabular}{ccccccccc}
\tablewidth{0pt}
\hline
\hline
Star No. & $\alpha$      &  $\delta$      & F606W     & $\delta$(F606W) & $A_{F606W}$ & F814W & $\delta$(F814W)  & $A_{F814W}$ \\
{}       & (deg J2000.0) &  (deg J2000.0) & (mag) & (mag)     & (mag)   & (mag) & (mag)     & (mag)   \\
\hline
0 & 247.79626 & 12.788096 & 19.30 & 0.001 & 0.15 & 18.32 & 0.001 & 0.10\\
1 & 247.79927 & 12.794552 & 19.22 & 0.001 & 0.15 & 18.30 & 0.001 & 0.09\\
2 & 247.78386 & 12.801684 & 19.32 & 0.001 & 0.15 & 18.44 & 0.001 & 0.09\\
\hline
\end{tabular}
\caption{Photometry of Hercules Center~2-ACS.} \label{tab:C2}
\begin{tabular}{ccccccccc}
\tablewidth{0pt}
\hline
\hline
Star No. & $\alpha$      &  $\delta$      & F606W     & $\delta$(F606W) & $A_{F606W}$ & F814W & $\delta$(F814W)  & $A_{F814W}$ \\
{}       & (deg J2000.0) &  (deg J2000.0) & (mag) & (mag)     & (mag)   & (mag) & (mag)     & (mag)   \\
\hline
0 & 247.72835 & 12.808475 & 19.54 & 0.001 & 0.15 & 18.56 & 0.001 & 0.09\\
1 & 247.73296 & 12.791601 & 19.49 & 0.001 & 0.15 & 18.84 & 0.001 & 0.10\\
2 & 247.73552 & 12.765498 & 20.67 & 0.001 & 0.16 & 18.43 & 0.002 & 0.10\\
\hline
\end{tabular}
\caption{Photometry of Hercules Center-WFC3.} \label{tab:CWFC3}
\begin{tabular}{ccccccccc}
\tablewidth{0pt}
\hline
\hline
Star No. & $\alpha$      &  $\delta$      & F606W     & $\delta$(F606W) & $A_{F606W}$ & F814W & $\delta$(F814W)  & $A_{F814W}$ \\
{}       & (deg J2000.0) &  (deg J2000.0) & (mag) & (mag)     & (mag)   & (mag) & (mag)     & (mag)   \\
\hline
0 & 247.85105 & 12.696190 & 19.47 & 0.001 & 0.16 & 18.29 & 0.001 & 0.10 \\
1 & 247.87623 & 12.688866 & 20.64 & 0.001 & 0.16 & 18.68 & 0.001 & 0.10 \\
2 & 247.84116 & 12.678297 & 20.06 & 0.001 & 0.16 & 19.28 & 0.001 & 0.10 \\
\hline
\end{tabular}
\caption{Photometry of Hercules West-ACS.} \label{tab:West}
\begin{tabular}{ccccccccc}
\tablewidth{0pt}
\hline
\hline
Star No. & $\alpha$      &  $\delta$      & F606W     & $\delta$(F606W) & $A_{F606W}$ & F814W & $\delta$(F814W)  & $A_{F814W}$ \\
{}       & (deg J2000.0) &  (deg J2000.0) & (mag) & (mag)     & (mag)   & (mag) & (mag)     & (mag)   \\
\hline
0 & 247.85105 & 12.696190 & 19.47 & 0.001 & 0.16 & 18.29 & 0.001 & 0.10 \\
1 & 247.87623 & 12.688866 & 20.64 & 0.001 & 0.16 & 18.68 & 0.001 & 0.10 \\
2 & 247.84116 & 12.678297 & 20.06 & 0.001 & 0.16 & 19.28 & 0.001 & 0.10 \\
\hline
\end{tabular}
\caption{Photometry of Hercules West-WFC3.} \label{tab:WWFC3}
\begin{tabular}{ccccccccc}
\tablewidth{0pt}
\hline
\hline
Star No. & $\alpha$      &  $\delta$      & F606W     & $\delta$(F606W) & $A_{F606W}$ & F814W & $\delta$(F814W)  & $A_{F814W}$ \\
{}       & (deg J2000.0) &  (deg J2000.0) & (mag) & (mag)     & (mag)   & (mag) & (mag)     & (mag)   \\
\hline
0 & 247.64271 & 12.794218 & 19.27 & 0.002 & 0.15 & 18.48 & 0.001 & 0.09 \\
1 & 247.66082 & 12.790000 & 19.80 & 0.001 & 0.15 & 18.15 & 0.001 & 0.09 \\
2 & 247.64324 & 12.805542 & 20.10 & 0.001 & 0.15 & 18.07 & 0.001 & 0.09 \\
\hline
\end{tabular}
\caption{Photometry of Hercules East-ACS.} \label{tab:East}
\begin{tabular}{ccccccccc}
\tablewidth{0pt}
\hline
\hline
Star No. & $\alpha$      &  $\delta$      & F606W     & $\delta$(F606W) & $A_{F606W}$ & F814W & $\delta$(F814W)  & $A_{F814W}$ \\
{}       & (deg J2000.0) &  (deg J2000.0) & (mag) & (mag)     & (mag)   & (mag) & (mag)     & (mag)   \\
\hline
0 & 247.95740 & 12.720261 & 19.38 & 0.001 & 0.14 & 18.41 & 0.001 & 0.09 \\
1 & 247.95326 & 12.750190 & 20.13 & 0.001 & 0.13 & 18.44 & 0.001 & 0.08 \\
2 & 247.93314 & 12.744528 & 20.24 & 0.001 & 0.14 & 18.42 & 0.001 & 0.08 \\
\hline
\end{tabular}
 \caption{Photometry of Hercules East-WFC3.} \label{tab:EWFC3}
\begin{tabular}{ccccccccc}
\tablewidth{0pt}
\hline
\hline
Star No. & $\alpha$      &  $\delta$      & F606W     & $\delta$(F606W) & $A_{F606W}$ & F814W & $\delta$(F814W)  & $A_{F814W}$ \\
{}       & (deg J2000.0) &  (deg J2000.0) & (mag) & (mag)     & (mag)   & (mag) & (mag)     & (mag)   \\
\hline
0 & 247.83546 & 12.802458 & 19.28 & 0.002 & 0.15 & 18.38 & 0.001 & 0.09 \\
1 & 247.87718 & 12.804322 & 19.66 & 0.001 & 0.14 & 18.60 & 0.001 & 0.09 \\
2 & 247.87247 & 12.784672 & 19.31 & 0.002 & 0.14 & 18.83 & 0.001 & 0.09 \\
\hline
\end{tabular}
   \begin{tablenotes}
      \small
      \item (These tables are available in theirs entirety in a machine-readable form in the online journal. A portion is shown here for guidance regarding its form and content.)
    \end{tablenotes}   
\end{minipage}    
\end{table*}

\begin{table}
\centering
\caption{Structural Properties of Hercules} \label{tab:str}
\begin{tabular}{lcc}
\tablewidth{0pt}
\hline
\hline
Parameter & Hercules & Ref. \\
\hline
R.A. (h m s)           & $16:31:03.00$     & 1 \\
Dec. (d m s)           & $+12:47:13.77$    & 1 \\
$M_{V}$ (mag)        & $-6.2\pm0.4$    & 1 \\     
$r_{h}$ (arcmin)     & $5.91\pm0.50$   & 1 \\     
$r_{h}$ (pc)   	     & $229.3\pm19.4$  & 1 \\  
Ellipticity          & $0.67\pm0.03$   & 1 \\ 
Position Angle (deg) & $-72.36\pm1.65$  & 1 \\
$m-M$ (mag)          & $20.58\pm0.02$  & 2 \\
Distance (kpc)       & $130.6\pm1.2$   & 2 \\  
$\langle E(B-V)\rangle$ &  0.06       & 2 \\
Heliocentric Velocity (km~s$^{-1}$) & $46.4\pm1.3$ & 3 \\
Velocity Dispersion (km~s$^{-1}$) & $4.4^{+1.4}_{-1.2}$ & 3 \\
$\mu_{\alpha} \cos{\delta}$$^{\star}$ (mas yr$^{-1}$) & $-0.153 \pm 0.074$ & 3 \\
$\mu_{\delta}$$^{\star}$ (mas yr$^{-1}$) & $-0.397 \pm 0.063$  & 3\\
Pericenter (kpc) & $50.9^{+24.2}_{-23.6}$ & 3 \\
Apocenter (kpc) & $227.9^{+85.1}_{-38.1}$ & 3 \\
Eccentricity & $0.65^{+0.10}_{-0.05}$ & 3 \\
\hline
\end{tabular}
  \begin{tablenotes}
      \small
      \item Notes: Last column is for references: (1) \citet{Sand2009}, (2) this work, (3) \citet{Gregory2020}. 
      \item $\star$ \citet{McConnachie2020} recently found a very similar systemic proper motion for Hercules by examining simultaneously the likelihood of the spatial, color-magnitude, and proper motion distribution of sources.
    \end{tablenotes}
\end{table}

\newpage
\section{COLOR MAGNITUDE DIAGRAMS}\label{sec:cmd}

\begin{figure*}
\centering
\includegraphics[width=1.7\columnwidth]{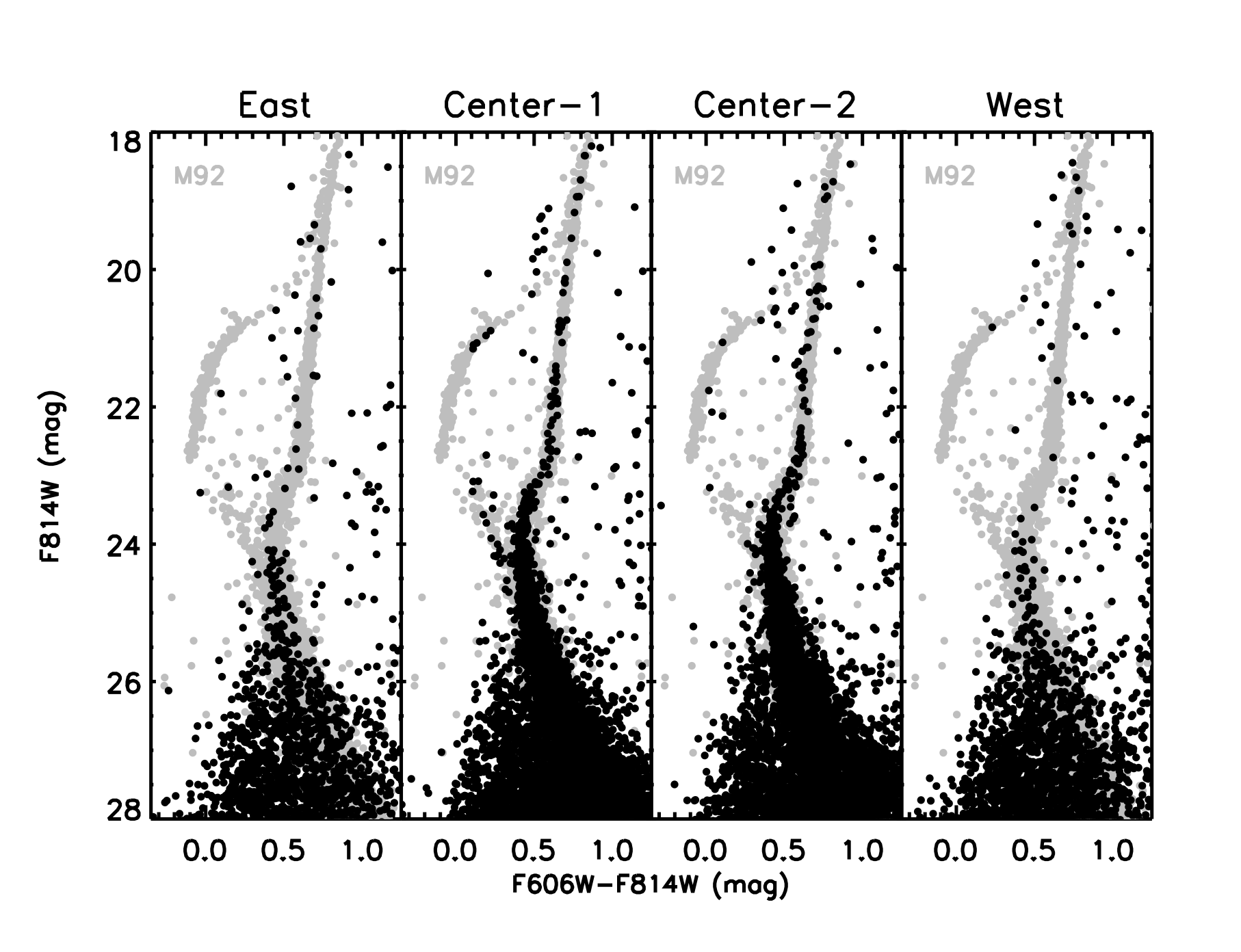}
\includegraphics[width=1.5\columnwidth]{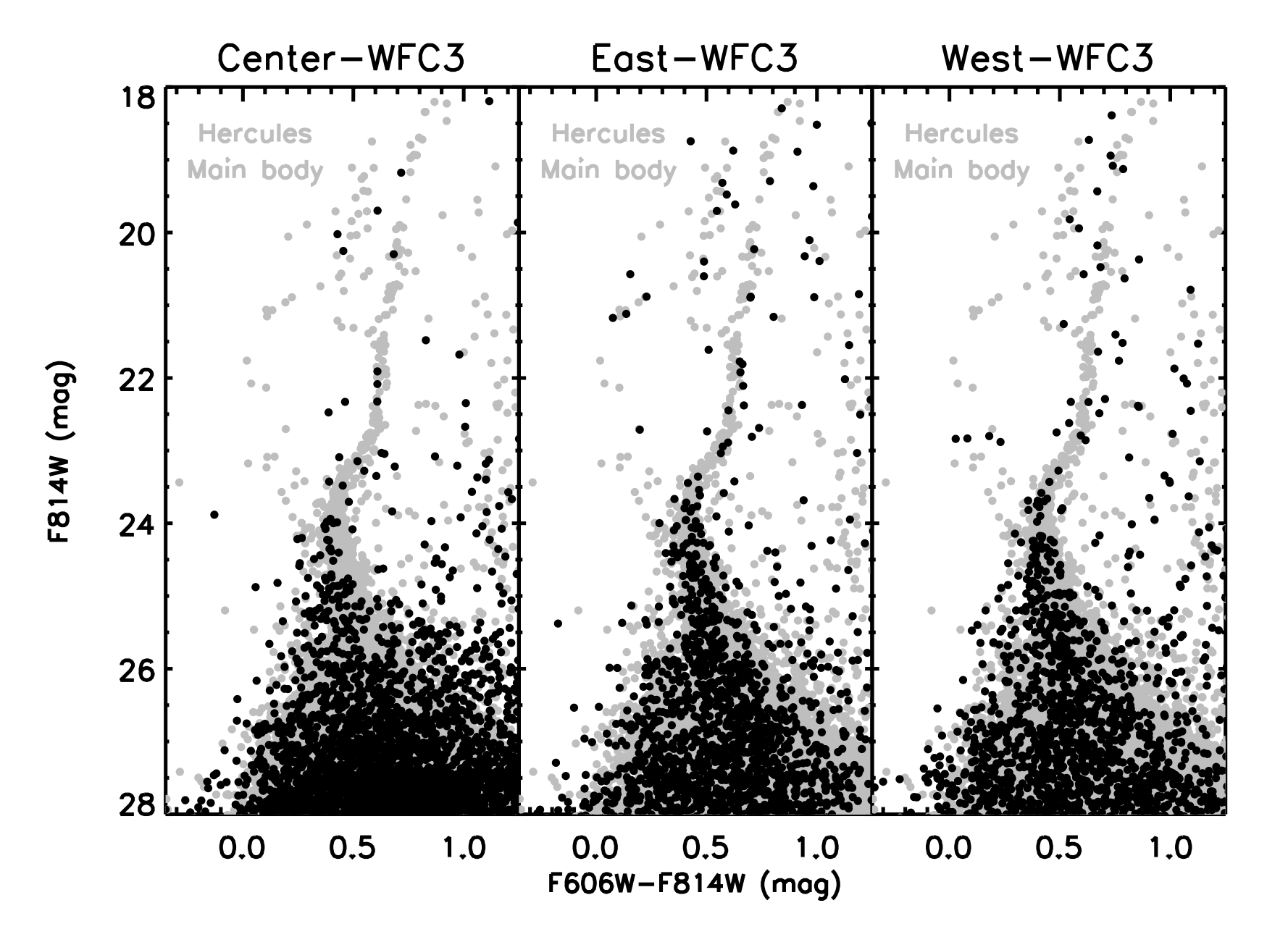}
\caption{CMDs of our Hercules {\it HST} fields (black points). The \textbf{top} panel shows the ACS fields relative to M92 (grey points). The M92 stars are shifted to the distance modulus of Hercules ($m-M=20.60$) from \citet{Sand2009}, for comparison purposes; M92 displays a very close agreement with Hercules, providing an important empirical fiducial for ancient metal-poor stellar populations. The off-center fields reveal a clear main sequence, supporting the stellar extended structure seen in ground-based observations is genuine \citep{Sand2009}. The \textbf{bottom} panel shows a comparison of the CMDs of the Hercules WFC3 fields, relative to the Hercules main body (i.e, Center-1$+$Center-2) shown in grey points.  \label{fig:figure2}}
\end{figure*}

Figure~\ref{fig:figure2} presents the CMDs of our Hercules {\it HST} fields, the top panel for the ACS fields relative to the Galactic globular cluster M92\footnote{The details of the M92 {\it HST} photometry and our derivation of its fiducial sequence were described in \citet{MutluPakdil2019}. Note that we implement the extinction correction for M92 using the same method described for Hercules, with an average $E(B-V)$ of 0.022 mag.} -- one of the most ancient, metal-poor, and well-studied star clusters known-- and the bottom panel for the WFC3 fields relative to the Hercules main body (i.e, Center-1$+$Center-2). 
There are a number of studies devoted to Hercules \citep[e.g.,][]{Coleman2007,Sand2009,Brown2014,Weisz2014,SG2007,Kirby2008,Koch2008,Aden2009}, agreeing that Hercules is old ($>12$~Gyr, with negligible star formation in the last 12~Gyr) and very metal-poor (with values of the mean metallicity $\langle[$Fe/H$]\rangle$ ranging from about $-2.0$ to $-2.7$~dex).
In Figure~\ref{fig:figure2}, the CMDs of the central Hercules fields have well-defined features with a clear main sequence turn-off (MSTO), displaying a close agreement with M92. A similarly close agreement with M92 was also found for Leo~V \citep{MutluPakdil2019}, which suggests that M92 is a nice fit to UFD galaxies that are dominated by ancient metal-poor populations. Note that metallicity estimates for M92 range from $-2.4 <$[Fe/H]$< -2.1$ \citep[e.g.,][]{Sneden2000,Behr2003,Carretta2009}, while there is evidence for [Fe/H]$<-2.5$ in individual M92 stars  \citep[e.g.,][]{Peterson1990,King1998,Roederer2011}. Therefore, M92 provides an important empirical fiducial for the stellar populations of Hercules, which we will use to revisit its distance in Section~\ref{sec:distance}.

The same central ACS fields were also studied by \citet{Brown2014}, along with five other ultra-faint dwarf galaxies (i.e., Bo\"{o}tes I, Canes Venatici II, Coma Berenices, Leo~V, and Ursa Major I). In addition to overall good agreement with M92, their CMDs show the presence of bluer and brighter stars near the MSTO when compared to the M92 ridge line, hence the authors suggested the presence of a very metal-poor stellar population in these ultra-faint dwarf galaxies. However, as we addressed in \citet{MutluPakdil2019}, the presence of a bluer and brighter star population in their CMDs can be explained with their adopted reddening values. \citeauthor{Brown2014} derived the distance and extinctions from fits to the ACS data and adopted $E(B - V) = 0.09$~mag for Hercules, which is higher than our adopted value ($E(B - V) = 0.06$~mag on average, see Table~\ref{tab:str}), which comes from the \citet{Schlafly2011} extinction derived from the \citet{Schlegel1998} reddening maps. 

In the top panel of Figure~\ref{fig:figure2}, the CMDs of the off-center East and West ACS fields, both located at $\sim2.5 \times$~half-light radii ($r_{h}$), reveal a clear main sequence, suggesting the elongated nature of  Hercules is indeed real and not clumps of compact background galaxies. We note that these stars still fall within $\sim2.5 \times r_{h}$, hence they might be bound stars. 
The bottom panel compares the CMDs of our WFC3 fields, relative to the central ACS fields. Note that East- and West-WFC3 are at similar projected radii, e.g., their average projected distance along the photometric major-axis ($d$) is $1.2$\arcmin and $2.0$\arcmin, respectively, while Center-WFC3 and our off-center ACS fields are at comparable projected radii (i.e., $d_{Center-WFC3}=7.0$\arcmin, $d_{East}=6.4$\arcmin, $d_{West}=7.5$\arcmin). In all of our off-center fields (i.e., East, West, Center/East/West-WFC3), the red giant branch (RGB) and subgiant branch (SGB) stars are not well populated, and the RGB suffers from significant field contamination, which is apparent from the scattering of stars beyond the M92 stellar locus. 
Although contamination includes unresolved background galaxies at magnitudes much fainter than the Hercules MSTO, it is dominated by Milky Way contaminants around the RGB. We will use the {\it Gaia}~DR2 data to disentangle a significant fraction of Milky Way contaminants from Hercules members in Section~\ref{sec:gaia}. Furthermore, there is a number of stars spread around a color of $\sim0.5$~mag and between F814W $\sim20.5$ to $\sim19$~mag, extending above the HB level (especially, in Center-1, there almost seems to be a parallel RGB), and we will also investigate their membership with {\it Gaia}~DR2 in the following section. 

\section{Membership Analysis}\label{sec:gaia}
\begin{figure*}
\centering
\includegraphics[width=1.\columnwidth]{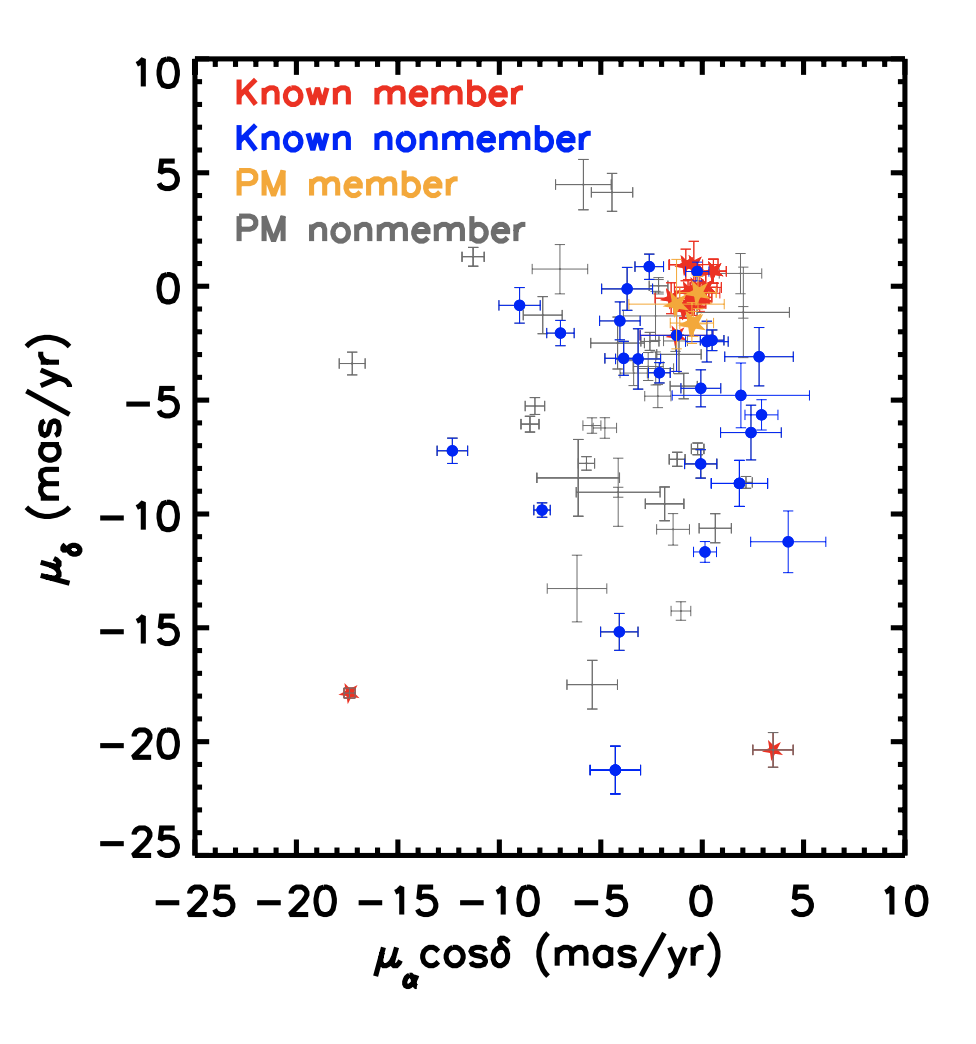}
\includegraphics[width=1.\columnwidth]{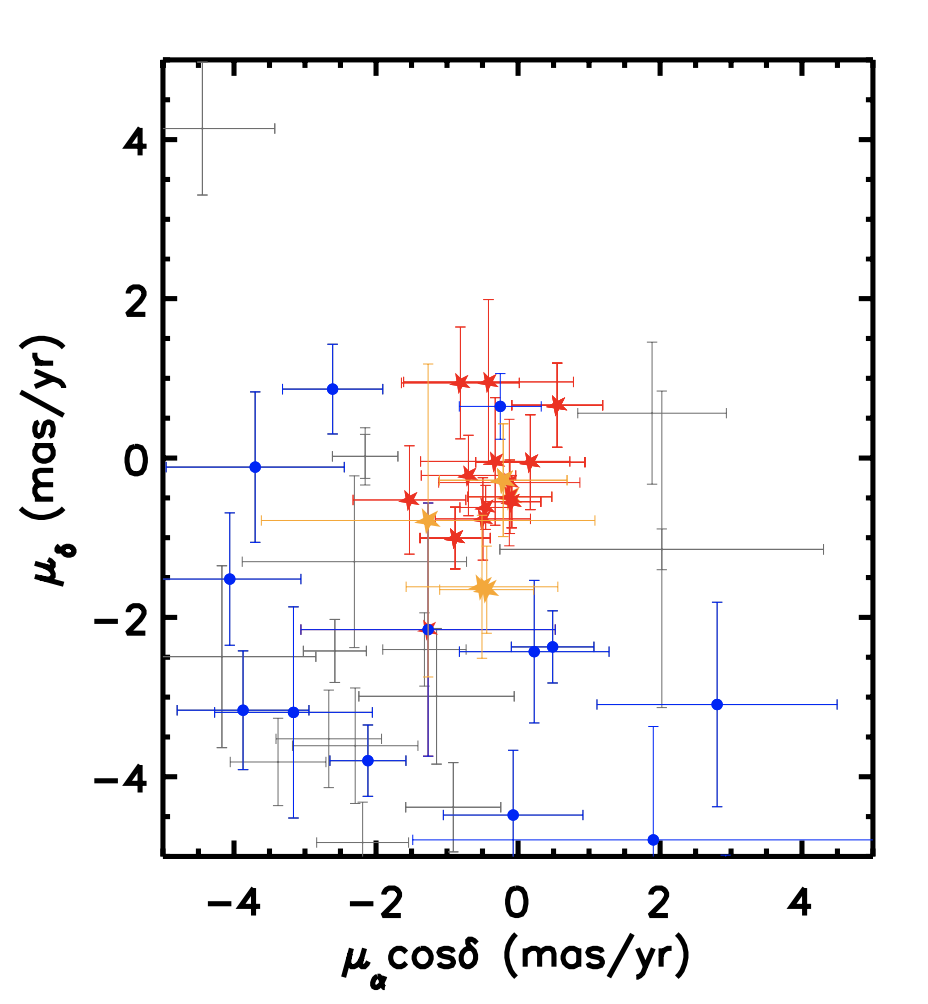}
\includegraphics[width=1.75\columnwidth]{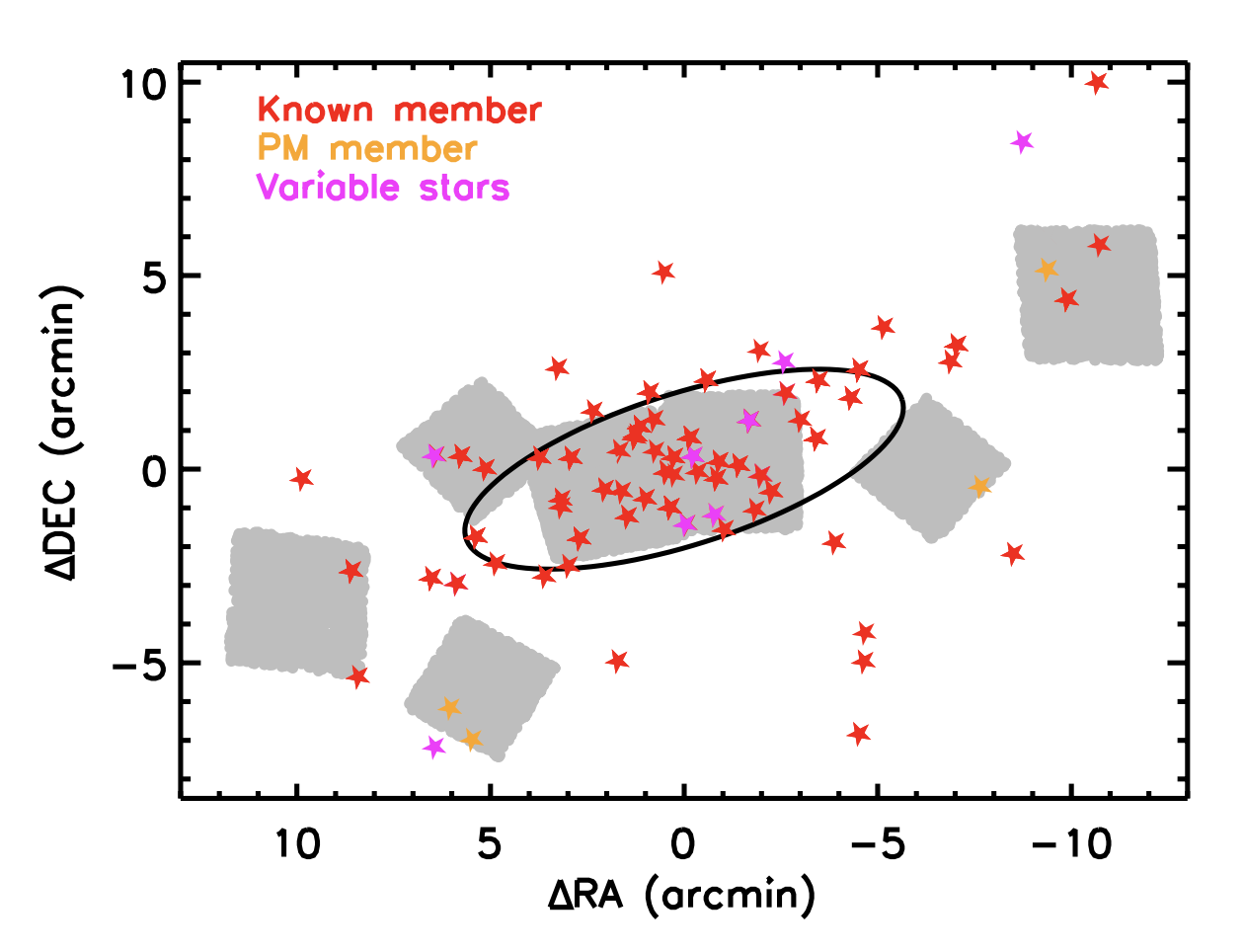}
\caption{\textbf{Top-Left:} Proper motions for all sources in our {\it HST} catalog with a match in {\it Gaia}~DR2, including the known kinematic members (red) and nonmembers (blue) identified in \citet{Aden2009,SG2007,Gregory2020}. \textbf{Top-Right:} Close up of the stars with acceptable proper motions. We identify stars without spectroscopy but with a proper motion consistent with those of know members as PM members.
\textbf{Bottom:} Spatial distribution of the Hercules kinematic members (red), variable stars (magenta), and our four new possible members (i.e., PM members, orange). We highlight the positions of our {\it HST} pointings with grey fields, and the ellipse marks the nominal half-light radius of Hercules. 
\label{fig:gaia}}
\end{figure*}

As shown in Figure~\ref{fig:figure2}, the outskirts of Hercules are poorly populated and heavily contaminated. Here, we review the known Hercules member and nonmember stars, and use {\it Gaia}~DR2 to explore the membership of bright stars (F814W $<20$~mag) in our {\it HST} fields. Our aim is to search for further Hercules member candidates while reducing the foreground contamination, according to their proper motions. First, we compile a comprehensive set of Hercules members, using membership catalogs from \citet{SG2007,Aden2009,Deason2012,Musella2012,Garling2018,Gregory2020}. 
Then, we cross-match our {\it HST} catalog to the {\it Gaia}~DR2 archive, selecting only those stars with a match within 1~arcsec. This catalogue matching returns a sample of 75 sources, of which 48 do not have any published spectroscopic data. The top-left panel of Figure~\ref{fig:gaia} shows the proper motions of all sources with a match in {\it Gaia}~DR2. The known radial velocity members (red stars) form a tight distribution in proper-motion space, except two with deviant proper motions -- one (located in Center-1) comes from the \citeauthor{SG2007} member catalog, the other one (located in West) comes from the \citeauthor{Gregory2020} catalog. We classify these two stars as PM nonmembers (PM stands for proper motion). In Table~\ref{tab:goodmembers}, we present a clean members catalog for Hercules, after refining their membership using {\it Gaia}~DR2 proper motions. Note that this catalog is based only on {\it HST} matches, so it does not extend beyond our {\it HST} footprint. As shown here (see also \citealt{Simon2018}), the stars that have been spectroscopically classified as UFD member stars in the literature deserve a second look as some of them might have {\it Gaia} proper motions very different from those of the galaxies which they supposedly belong to. 

\begin{figure*}[!hp]
\centering
\includegraphics[width=1\linewidth]{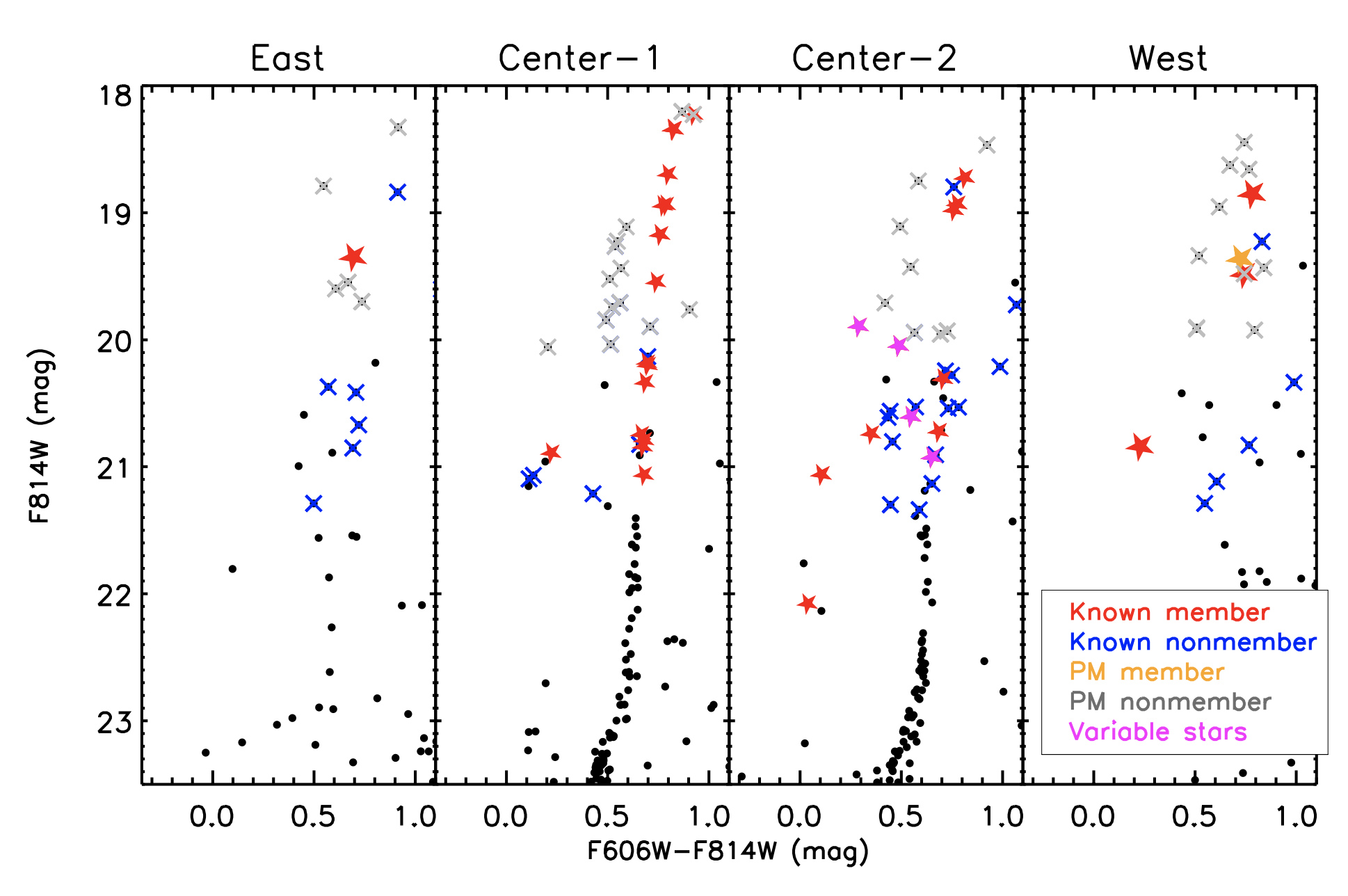}
\includegraphics[width=0.75\linewidth]{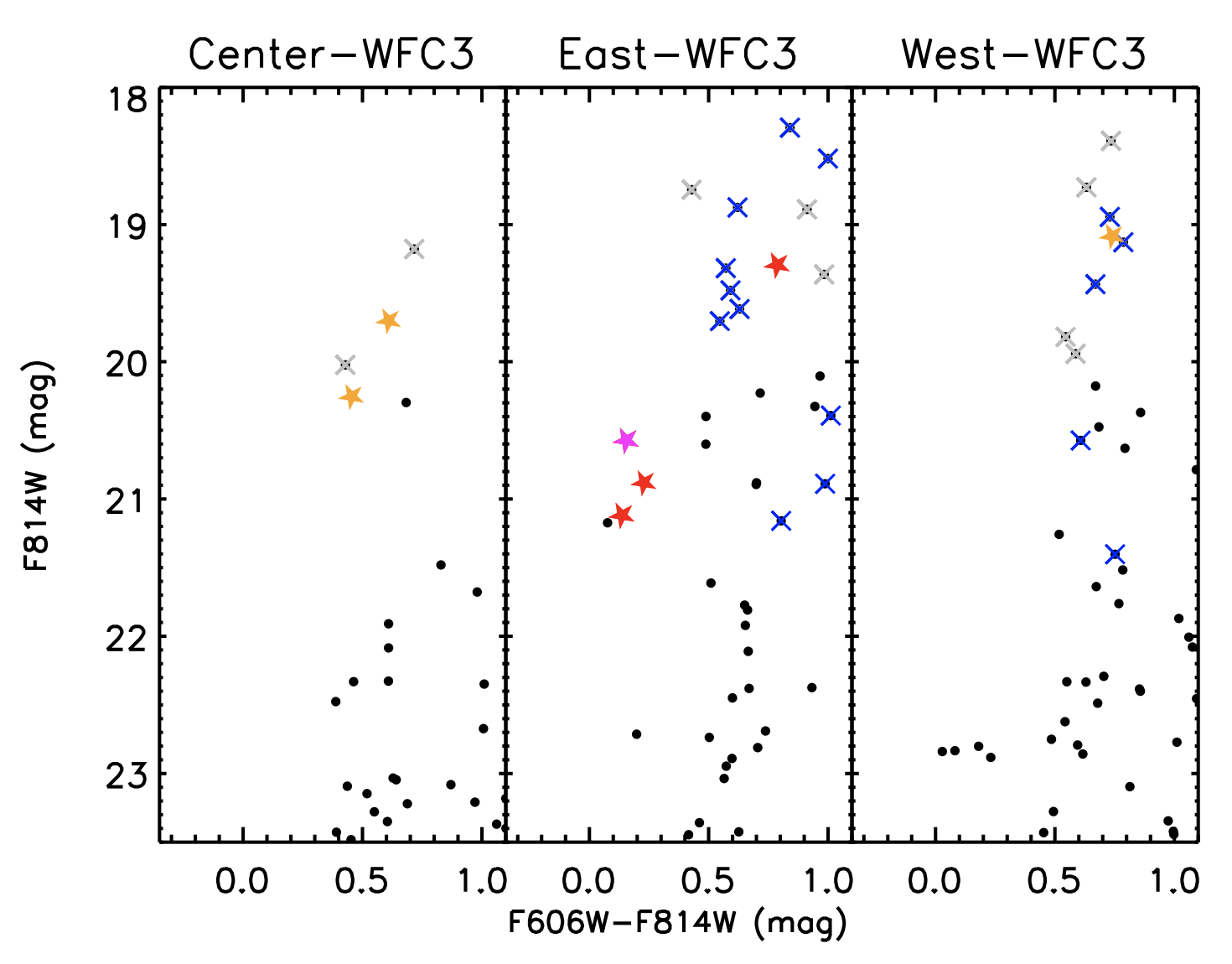}
\caption{CMDs of our {\it HST} fields, highlighting the stars in Figure~\ref{fig:gaia}. Our proper motion analysis allows us to disentangle a significant fraction of field contamination in the vicinity of the RGB (see PM nonmembers -- grey crosses), while providing four new member candidates (see PM members -- orange stars). \label{fig:cmdmembers}}
\end{figure*}

The top-right panel of Figure~\ref{fig:gaia} shows a close up of the stars with acceptable proper motions, which we define as $\vert (\mu_{\alpha} \cos{\delta},\mu_{\delta}) \vert \leq 2$ mas yr$^{-1}$. There are four new stars without spectroscopy but with a proper motion in this accepted range, and we label them as PM members in the figures and in Table~\ref{tab:maybemembers}. These stars are ideal targets for future spectroscopic studies of Hercules. We classify the rest of the stars without spectroscopy as PM nonmember, as their proper motions imply that they are likely not a true Hercules member. The bottom panel of Figure~\ref{fig:gaia} shows the spatial distribution of our new PM members, along with the known members (each defined with a particular legend). We show the positions of these stars in the color-magnitude space in Figure~\ref{fig:cmdmembers}. Our PM members located in the West and West-WFC3 fields are well within the RGB locus while the remaining two, located in Center-WFC3, are good RR Lyrae candidates. Our PM nonmembers help us to reduce heavy field contamination in the vicinity of the RGB. Stars with a color of $\sim0.5$~mag and magnitudes in the range $18.0\lesssim$ F814W $\lesssim 20.5$~mag, extending above the HB level, were suggested to be variable stars \citep{Aden2009}, however our {\it Gaia} investigation reveals that most are PM nonmembers and thus are not associated with Hercules.  

In short, using the {\it Gaia}-DR2 data, we clean the literature spectroscopic samples from nonmembers and compile a set of robustly identified Hercules members while providing a new target list for further spectroscopic observations. We use this information to update the properties of Hercules (e.g., distance) and search for any signs of a distance gradient in this data.

\begin{table*}
\centering
\small
\caption{A clean members catalog for Hercules within our {\it HST} footprint}\label{tab:goodmembers}
\begin{tabular}{lcccccccc}
\tablewidth{0pt}
\hline
\hline
{No} & R.A. & Dec & F606W & F814W & $\mu_{\alpha} \cos{\delta}$ & $\mu_{\delta}$ & {Field} & {Reference}\\
{} & (deg) & (deg) & (mag) & (mag) & (mas yr$^{-1}$) & (mas yr$^{-1}$)  &  {} & {}\\
\hline
1 &	247.80862  &  12.757386   & 19.17  &     18.34 & $-0.089\pm0.410$ & $-0.548\pm0.328$ & Center-1 & (1,2)\\
2 &	247.78386  &  12.801684   & 19.17  &     18.34 & $-0.457\pm0.362$ & $-0.620\pm0.276$ & Center-1 & (1,2,6)\\
3 &	247.77063  &  12.785917   & 19.49  &     18.70 & $-0.885\pm0.495$ & $-1.004\pm0.387$ & Center-1 & (1)\\
\hline
\end{tabular}
 \begin{tablenotes}
      \small
      \item NOTE $-$ Column~1 lists our assigned number for each star. Columns~2-5 are the right ascension, declination, F606W and F814W magnitudes from our {\it HST} catalog (if they are within our fields), respectively. Columns~6-7 are the \textit{Gaia} DR2 proper motions. Column~8 lists other IDs for each star from the literature. References are listed in Column~9: (1) \citet{SG2007}, (2) \citet{Aden2009} (3) \citet{Deason2012}, (4) \citet{Musella2012}, (5) \citet{Garling2018}, (6) \citet{Gregory2020}
      \item \textit{(This table is available in its entirety in a machine-readable form in the online journal. A portion is shown here for guidance regarding its form and content.)}
\end{tablenotes} 
\end{table*}

\begin{table*}
\centering
\small
\caption{List of plausible Hercules members, requiring further follow-up to confirm.}\label{tab:maybemembers}
\begin{tabular}{lccccccc}
\tablewidth{0pt}
\hline
\hline
{Field} & R.A. & Dec & F606W & F814W & $\mu_{\alpha} \cos{\delta}$ & $\mu_{\delta}$ & {CMD}\\
{} & (deg) & (deg) & (mag) & (mag) & (mas yr$^{-1}$) & (mas yr$^{-1}$)  &  {} \\
\hline
West & 247.60223 & 12.873091 & 20.09 & 19.36 & $-0.209\pm0.899$ & $-0.278\pm0.708$ & RGB \\
West-WFC3 & 247.63171 & 12.779719 & 19.82 & 19.08 & $-0.441\pm0.662$ & $-1.652\pm0.547$ & RGB \\
Center-WFC3 & 247.85607 & 12.67086  & 20.31 & 19.70 & $-0.508\pm1.068$ & $-1.616\pm0.897$ & RRL\\
Center-WFC3 & 247.86572 & 12.684449 & 20.71 & 20.25 & $-1.266\pm2.349$ & $-0.784\pm1.964$ & RRL\\
\hline
\end{tabular}
  \begin{tablenotes}
  \small
      \item NOTE $-$ Columns~1-5 are the field name, the right ascension, declination, F606W and F814W magnitudes from our {\it HST} photometry, respectively. Columns~6-7 are the \textit{Gaia}~DR2 proper motions. Column~8 reflects the position on CMD. 
\end{tablenotes}
\end{table*}

\newpage
\section{EXPLORING THE DISTANCE GRADIENT ACROSS HERCULES}\label{sec:distance}

Our goal is to explore whether or not Hercules presents a significant distance gradient across its projected length on the sky. We refer the reader to Figure~\ref{fig:figure1} for our observational strategy. We also remind them that Center-1 and -2 are centered on the main body of Hercules while our other fields trace the outskirts of the galaxy. As shown in Section~\ref{sec:cmd}, M92 provides an important empirical fiducial for the stellar populations of Hercules, therefore the properties of Hercules can be robustly measured by making a comparison to the ridgeline of M92. Finally, we note that our {\it HST} fields reach a similar depth (see Table~\ref{tab:obslog}), hence they are comparable.

\begin{figure*}
\centering
\includegraphics[width=0.63\linewidth]{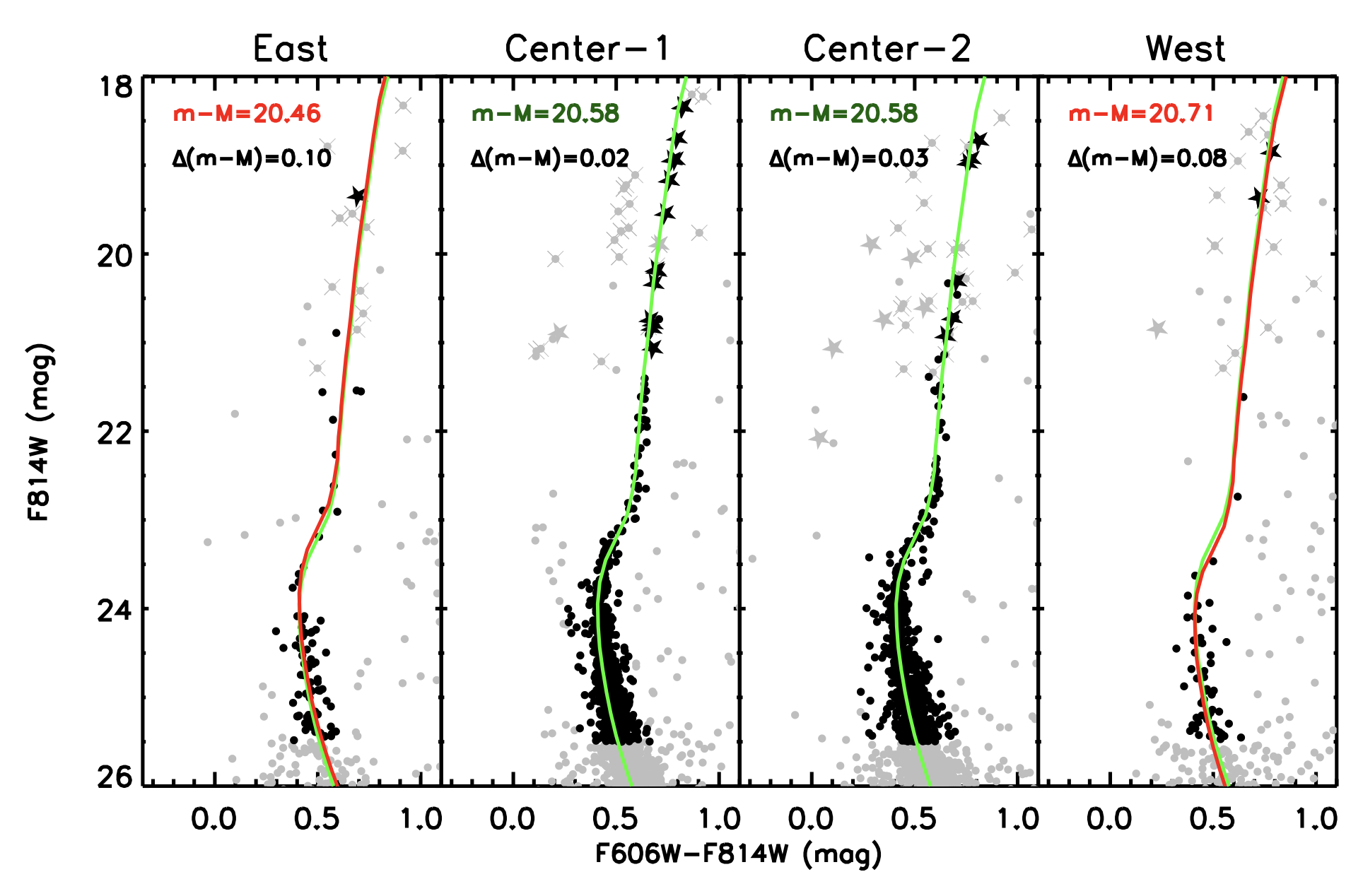}
\includegraphics[width=0.51\linewidth]{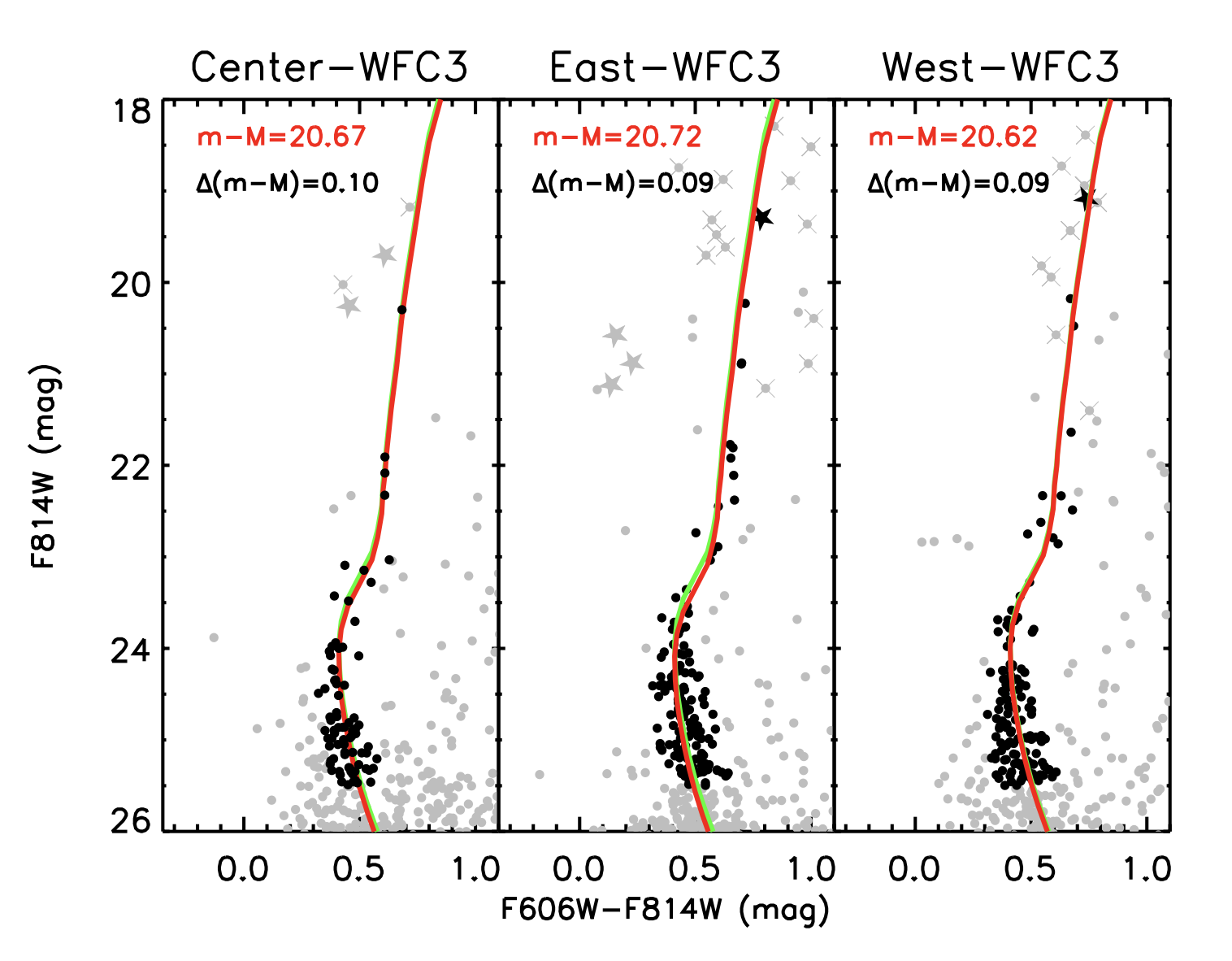}
\includegraphics[width=0.5\linewidth]{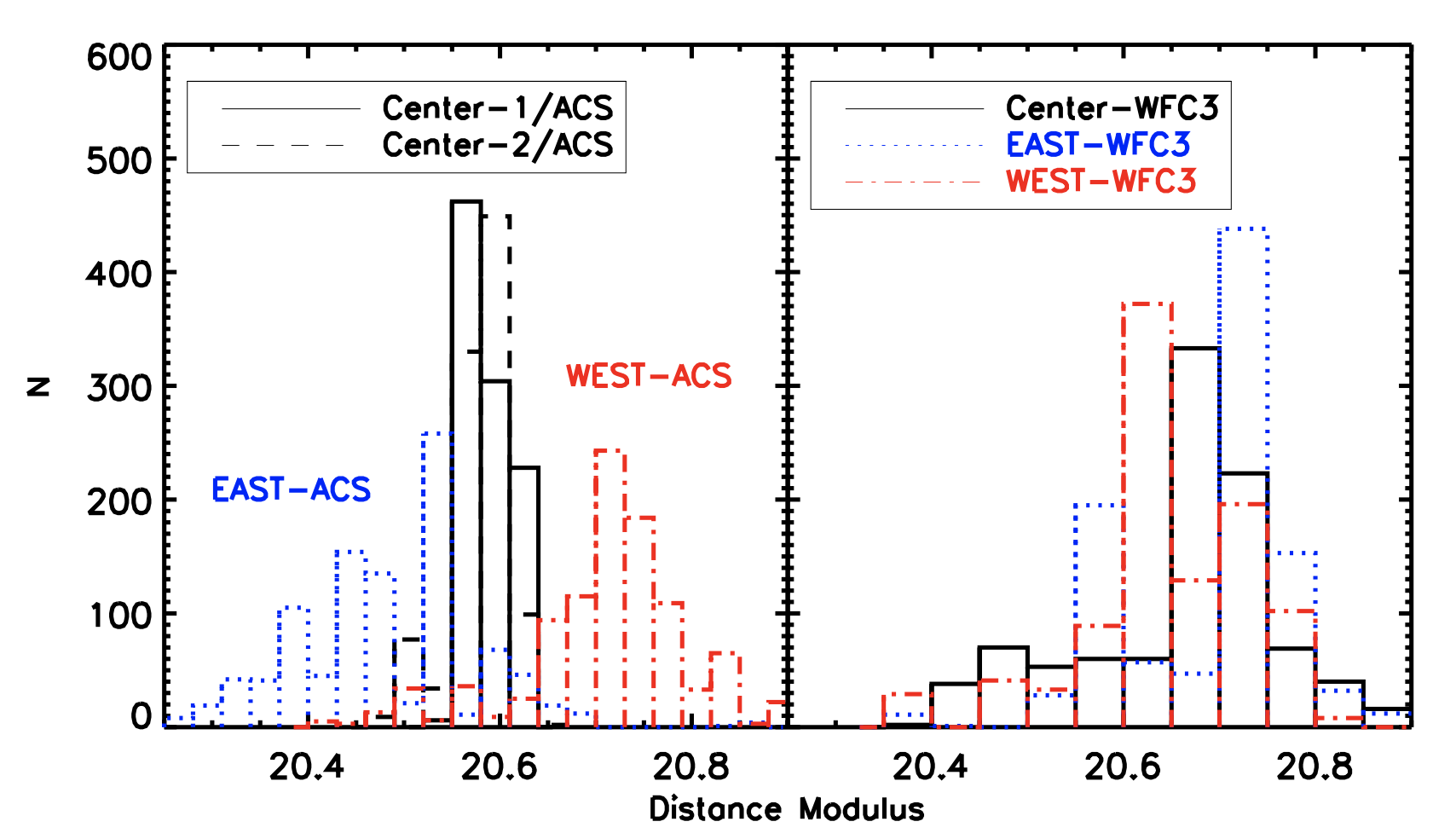}
\includegraphics[width=0.41\linewidth]{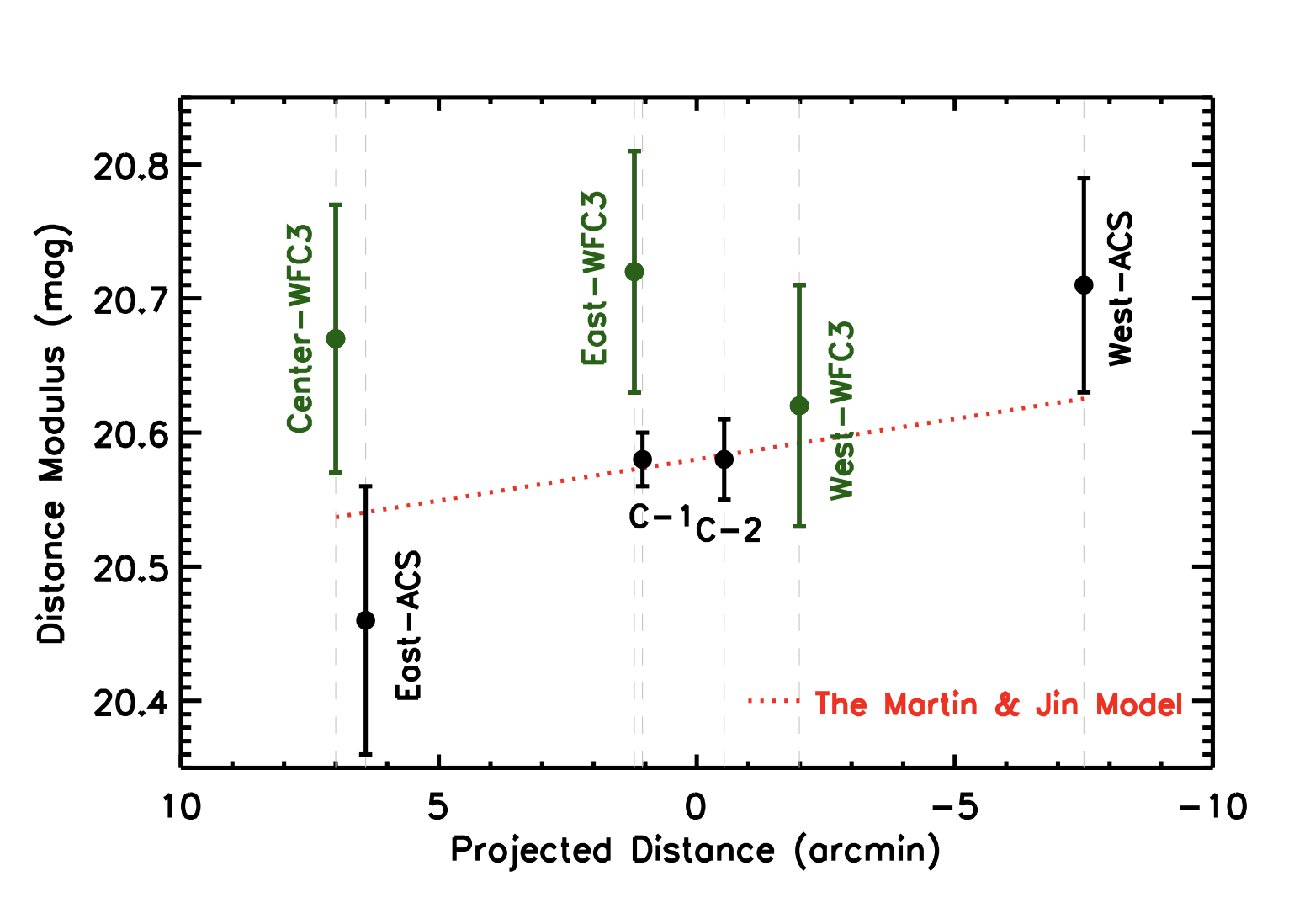}
\caption{\textbf{Top/Middle:} CMDs of the stars in our ACS/WFC3 fields (grey points), highlighting the ones used in our distance analysis (black points). Hercules' likely members -- radial velocity and PM members -- are shown as filled stars. Overplotted as a red line is the M92 fiducial at our derived distance for each individual field, and the green line refers to the one at $m-M=20.58$~mag (the common value derived for the central two fields), shown for comparison purposes in all panels. $\Delta (m-M)$ is the uncertainty derived from our bootstrapping analysis. \textbf{Bottom-left/middle:} Bootstrap histograms of the distance modulus for our ACS/WFC3 fields. \textbf{Bottom-right:} Distance modulus versus the projected distance of each field. For each field, we adopt the median of our bootstrap realizations as our final result, and use its standard deviation as our final uncertainty. Note that East-ACS seems to be closer (123.6~kpc) than West-ACS, however the WFC3 fields weaken this distance gradient argument. 
\label{fig:distance}}
\end{figure*}

As a first test, we measure the distance modulus of each individual field with a simple chi-squared minimization routine based on the differences between the shape of the M92 fiducial and the observed sequences. We assume a distance modulus of $m - M = 14.62$~mag for M92 as in \citet{Brown2014}, taking the mean of the measurements from \citet[$14.60\pm0.09$ mag]{Paust2007}, \citet[$14.62\pm0.1$ mag]{DelPrincipe2005}, and \citet[$14.65\pm0.1$ mag]{Sollima2006}. To reduce the contamination by foreground stars, we construct a selection region covering the RGB to SGB on our CMDs -- stars brighter than 25.5~mag in F814W. First, we perform a CMD-selection by including sources with colors and magnitudes expected for the stellar population of Hercules. A fairly broad selection is chosen here to ensure that we do not reject a significant fraction of the dwarf members. More specifically, we inflate the uncertainty to $0.05$, $0.08$, and $0.1$~mag for F814W$< 21$, $21 \leq$ F814W $< 24$, F814W$\ge24$~mag, respectively. Note that our photometric errors are much smaller than these floor uncertainties that we adopt. The selected stars are later cleaned of known nonmembers and stars whose {\it Gaia} proper motions are not consistent with those of the Hercules kinematic members (see Section~\ref{sec:gaia}). The final star catalogs used in our distance analysis are shown in Figure~\ref{fig:distance}.  

The M92 fiducial is shifted through $0.01$~mag intervals in $(m - M)$ from $20.20$ to $21.0$~mag in F814W, a plausible range of distance moduli for Hercules. In each step, we calculate the chi-square statistic as
\begin{equation}\label{eqn:1} 
\chi^2 = \sum\limits \frac{(c_{i,O}-c_{i,E})^2}{\sigma({c_{i,O}})^2}
\end{equation}
where $c_{i,O}$ is the observed color of the i-th star, $c_{i,E}$ is the expected color of the same star on the fiducial at its magnitude, and $\sigma({c_{i,O}})$ is its photometric uncertainty in color (based on our artificial star tests). In this step, we impose a floor photometric uncertainty of $0.005$~mag, but we note that our true photometric uncertainties are smaller at F814W$\lesssim22.0$~mag. We adopt the distance modulus that minimizes the chi-square statistic. Within our clean catalog, there might be stars associated with Hercules -- either currently bound or in tidal material -- and field stars. We use a 1000 iteration bootstrap analysis to determine the uncertainties due to the remaining contaminants. Especially for the off-center fields, where the number of stars is much smaller, the field star contamination might significantly affect distance measurements, and it is critical to properly take into account the known Hercules members in these poorly populated fields. Therefore, in each resampling, we purposely keep the known Hercules members and draw randomly from the remaining stars. From the 1000 realizations, we take the median as our final distance measurement and its standard deviation as our uncertainty.

We present the results of our bootstrap analysis in Figure~\ref{fig:distance}, along with CMDs of our fields overlaid with the M92 fiducial at our derived distances. The bootstrap histograms of the central ACS fields are well-defined with an overlapping median at $20.58$~mag and a small standard deviation ($\sim0.02$~mag). This translates to $130.6\pm1.2$~kpc, which is consistent with the distance estimates in the literature ranging from $132\pm6$~kpc to $147^{+8}_{-7}$~kpc \citep[e.g.,][]{Belokurov2007,Coleman2007,Aden2009,Sand2009,Musella2012,Garling2018}. In particular, our measurement is in perfect agreement with the distance estimations from RR Lyrae stars ($132\pm6$~kpc, \citealt{Musella2012}; $137\pm11$~kpc, \citealt{Garling2018}). It should be noted that our distance uncertainty is only associated with the fit to our photometry, and do not include systematic errors associated with the M92 distance (the distance to M92 is uncertain at the level of $\lesssim0.1$~mag). Since the off-center ACS fields are poorly populated, their histograms are broader with a standard deviation of $\sim0.1$~mag. Intriguingly, the eastern bootstrap histogram has a median of $(m-M)=20.46$~mag (123.6~kpc) while the western one peaks around $(m-M)=20.71$~mag (138.7~kpc), implying a distance gradient of $\sim0.25$~mag ($\sim15$~kpc) across these off-center ACS fields. While this is interesting, the bootstrap histograms of the WFC3 fields (Figure~\ref{fig:distance}, bottom-middle) have a significant overlap with an almost opposite trend (i.e., West-WFC3 seems to be slightly closer than East-WFC3 as its bootstrap histogram peaks at a lower distance modulus), casting doubt on the existence of a significant gradient. This is more clearly illustrated in the bottom-right panel, which shows the distance modulus of each field as a function of the projected galactocentric distance. If the gradient is as strong as the ACS-only fields, then new intermediate ACS fields, which would have lower uncertainties, might provide significantly stronger constraints on the gradient.

\begin{figure}
\centering
\includegraphics[width=0.95\linewidth]{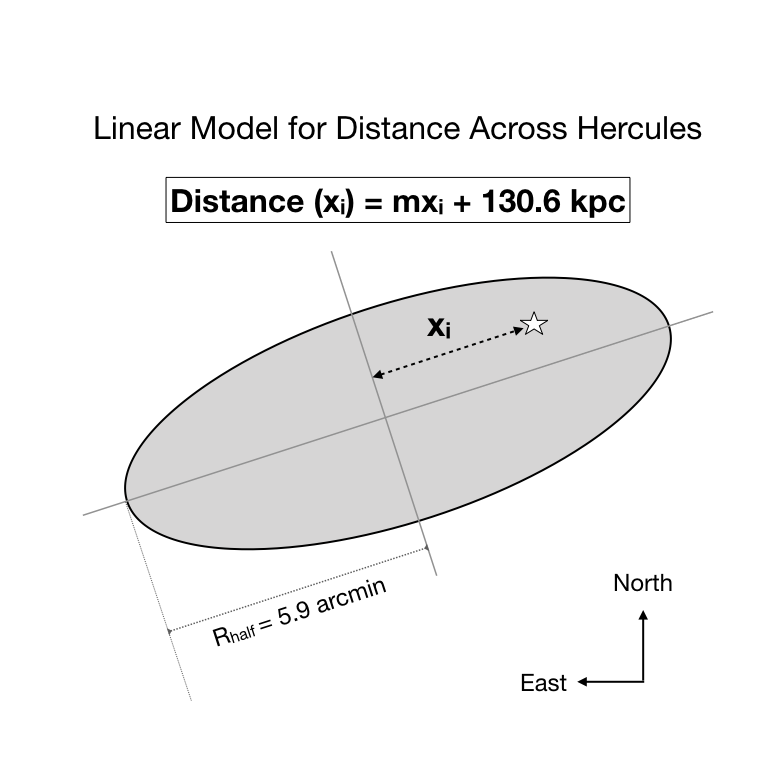}
\includegraphics[width=0.95\linewidth]{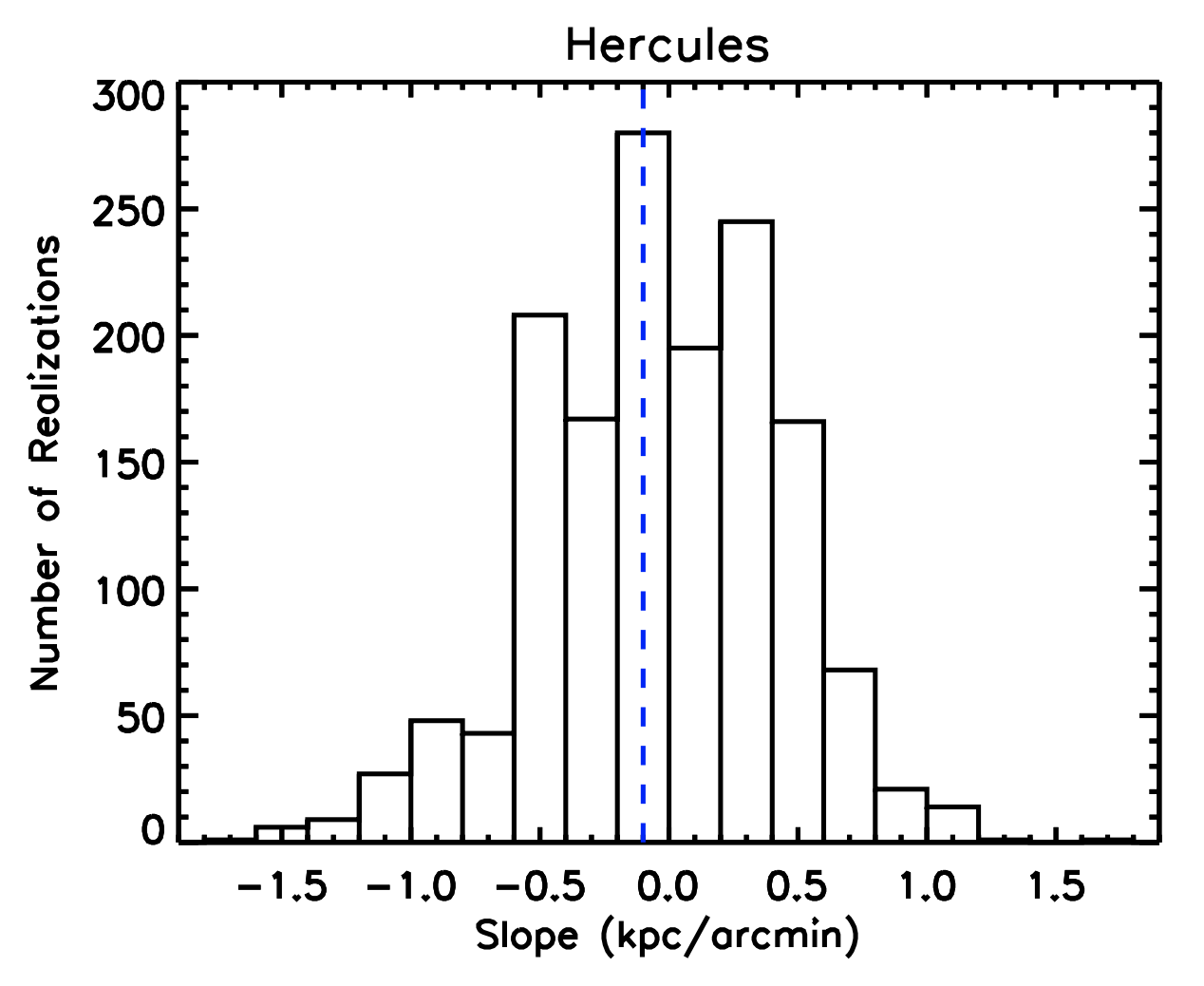}
\caption{\textbf{Top:} Illustration of our model for the observer–Hercules distance changing as a function of the major-axis distance. \textbf{Bottom:} Histogram of the best-fitting slope with 1500 bootstrap resamples for which we fit our linearly changing distance model. The dashed line corresponds to the median, which is consistent with a zero or negligible slope. 
\label{fig:slope}}
\end{figure}

To further explore the distance gradient, we fit a model with Hercules' distance changing linearly as a function of the major-axis distance, as previously done in \citet{Sand2009}. This model assumes that Hercules is no longer a bound dwarf galaxy but instead a stellar overdensity in a thin stream whose length we are observing nearly along the line-of-sight. We adopt $(m-M)=20.58$~mag (130.6~kpc) for the center of Hercules (see Figure~\ref{fig:distance}), and then allow the observed-Hercules distance to change as a function of the major-axis distance: 
\begin{equation}\label{eqn:2}  
\mbox{Distance} (x_{i}) = mx_{i} + 130.6~\mbox{(kpc)} 
\end{equation}
For a given slope, $m$, and major-axis distance for the i-th star, $x_i$, the presumed distance to a Hercules member is known, and the M92 fiducial is then transformed to that distance to interpolate the expected color. We choose to vary $m$ between $-2.5$ and $3.6$~kpc/arcmin, a plausible range of distance gradient for Hercules, with $0.05$~kpc/arcmin intervals. We note that the slope proposed by \citet{MartinJin2010} is $-0.37$~kpc/arcmin, which is well within the range we explore here. In each step, we calculate the chi-square statistic, as in Equation~\ref{eqn:1}. The cartoon in Figure~\ref{fig:slope} illustrates our model, and the bottom panel shows the results over 1500 bootstrap resamples. The bootstrap-derived histogram peaks at the slope of $-0.10$~kpc/arcmin, which is consistent with a zero or negligible slope, indicating no measurable distance gradient across the face of Hercules. However, the histogram is broad with a standard deviation of $0.48$~kpc/arcmin, which still makes the \citeauthor{MartinJin2010} model plausible.  

\subsection{Simulated Datasets}
Before moving forward, we explore how well our method recovers the truth within the statistical uncertainty. We also want to better understand the limitations of our methodology. A natural step is to apply our two tests (one is deriving the distance for individual fields, the other is fitting a linearly-changing distance model) on a series of artificial Hercules analogs with known distance gradients. We construct our model galaxy catalogs by sampling random stars from our M92 fiducial (after accounting for our photometric errors and adopting the observed luminosity function of Hercules), and placing stars using an exponential profile with the structural parameters (half-light radius$=5.9$\arcmin, ellipticity$=0.67$, position angle$=-72^{\circ}$) in Table~\ref{tab:str}. We focus on four categories: 1) no distance gradient, 2) a slope of -0.10~kpc/arcmin, 3) the slope of the \citeauthor{MartinJin2010} model (i.e., $-0.37$~kpc/arcmin), and 4) a slope of -0.50~kpc/arcmin. Then, each star is shifted to the distance predicted by Equation~\ref{eqn:2}. A total of 50 galaxies are generated in each category.   

We utilize just the artificial stars located at the corresponding distances of our {\it HST} fields, treat these simulated star catalogs in the same way as our real data, and perform our tests as mentioned above. Overall, our first method is very successful at recovering the true mean distance of each individual field within the estimated uncertainties. However, in the cases of a significant distance gradient (Categories~3 and 4), we also find a number of simulated dwarf galaxies whose off-center fields show a flat or even opposite distance trend (just like in our WFC3 fields, see Figure~\ref{fig:distance}-bottom). Similarly, it is possible to derive a strong gradient in some of the model galaxies with an intrinsically negligible distance gradient (Categories~1 and 2) due to ‘‘CMD shot noise’’ as  described in \citet{Martin2008}. A key point here is that the result must be taken with caution -- while not finding a clear difference between the eastern and western portions of the galaxy reduces the likelihood of a real significant distance gradient, finding a flat trend between these sparsely-populated off-center fields does not mean that Hercules has no distance gradient along the line-of-sight. 

\begin{figure}
\centering
\includegraphics[width=0.97\linewidth]{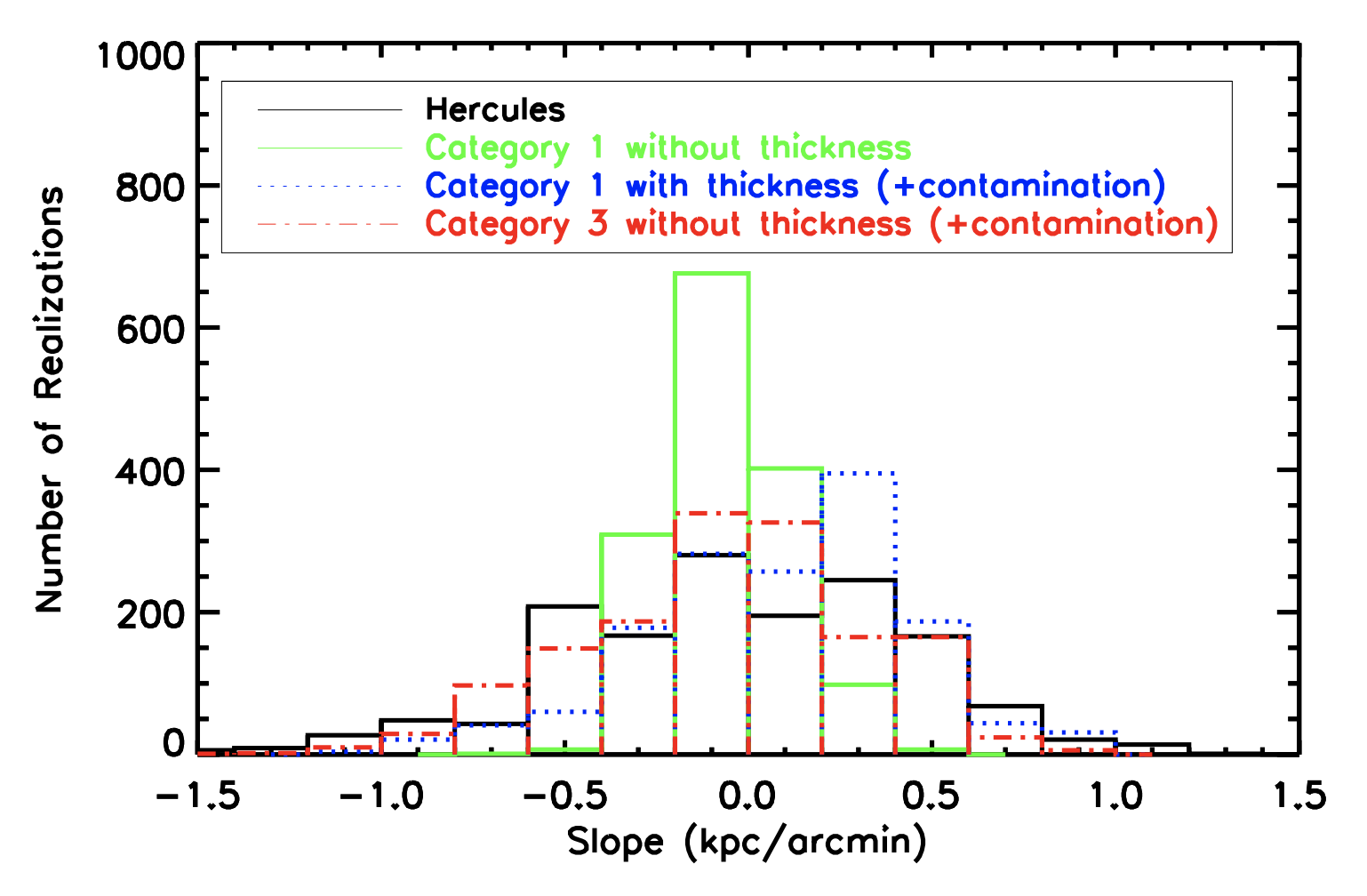}
\caption{The histograms of the best-fitting slope for individual simulated dwarfs relative to that of Hercules. The green line shows an example Category~1 model galaxy (no distance slope, the estimated slope is $-0.10\pm0.18$~kpc/arcmin). The other two examples include additional field contamination: the blue dotted-line refers to an example Category~1 simulation with a thickness of $\sim1$~kpc (the estimated slope is $0.15\pm0.37$~kpc/arcmin) while the red dash-dotted line shows an example Category~3 simulation (the \citeauthor{MartinJin2010} model) without an intrinsic thickness component (the estimated slope is  $-0.10\pm0.40$~kpc/arcmin). Note that we find a slope of $-0.10\pm0.48$~kpc/arcmin for our real Hercules data. Both the thickness of the dwarf and field contamination introduce additional uncertainties, which makes it harder to distinguish between different models.  
\label{fig:slope2}}
\end{figure}

Next, we explore our star-by-star slope-fitting method on the simulated galaxies, and find that their bootstrap-derived histograms have a shape well represented by a Gaussian approximately centered on the true slope. Interestingly, these histograms are much narrower than the one we get for Hercules with an average scatter of $\sim0.2$~kpc/arcmin (see the green line in Figure~\ref{fig:slope2}). 
To explore the impact of field contamination, we use TRILEGAL, a simulator of photometry for stellar populations in the Galaxy \citep{Girardi2012}, and estimate the number of Galactic sources within the appropriate area. Considering that we already remove the nonmember stars while estimating the distance, we expect $\sim2-6$ more field stars per field. 
If we add field contamination to the model galaxies, their histograms become broader. However, their median might deviate from the true slope, as shown with an example Category~3 galaxy (with the slope of $-0.37$~kpc/arcmin, i.e, the \citeauthor{MartinJin2010} model), whose histogram peaks at $-0.10$~kpc/arcmin with a scatter of $0.40$~kpc/arcmin in the presence of field contamination (see the red dash-dotted line).
To better probe the complex nature of Hercules, we perform an additional test including an intrinsic thickness component to the stream, where each star is additionally shifted to a distance randomly sampled from a Normal distribution with a standard deviation of $\sim1$~kpc (note that $r_h=0.2$~kpc for Hercules, see Table~\ref{tab:str}). The thickness of a dwarf galaxy introduces additional uncertainty, and we find that on average the slope is overestimated half of the time. Overall, this means that constraining a distance gradient in such a faint system is not trivial, and the thickness of the dwarf and field contamination make it harder to distinguish different models.
  
\section{Conclusions}\label{sec:conclusion}

In this work, we present a comprehensive deep {\it HST} imaging study of Hercules and its surrounding regions, combined with the {\it Gaia}~DR2 archival data, and perform a new observational test for its orbital models by constraining the presence of a distance gradient in order to understand the peculiar properties of this ultra-faint system. Here, we summarize our key results:

\begin{itemize}
    \item For a better understanding of the properties of Hercules, we make a comparison with the Galactic globular cluster M92. The CMDs of the central Hercules fields display a close agreement with that of M92, implying they have similar stellar populations and star formation histories. 
    \item The CMDs of our off-center fields reveal a clear main sequence, supporting  the idea that the stellar extension seen along the major-axis of Hercules is real, not a clump of background galaxies \citep[as found in Leo~V,][]{MutluPakdil2019}.
    \item As expected, our off-center fields are poorly populated and their RGBs suffer from significant field contamination. We utilize the {\it Gaia}~DR2 proper motion data to disentangle a significant fraction of field contaminants from Hercules members. We clean the literature spectroscopic samples from nonmembers (e.g., two previously identified spectroscopic members are now classified as PM nonmembers), compile a set of robustly identified Hercules members (see Table~\ref{tab:goodmembers}), and provide a new target list for further spectroscopic observations (i.e., two RGB and two RR Lyrae candidates which we classify as PM members, see Table~\ref{tab:maybemembers}). 
    \item We update the distance of Hercules with isochrone fitting, and find a distance of $130.6\pm1.2$~kpc ($m-M=20.58\pm0.02$) for the main body, which is in excellent agreement with the distance measurements from RR Lyrae stars ($132\pm6$~kpc, \citealt{Musella2012}; $137\pm11$~kpc, \citealt{Garling2018}). 
    \item Leveraging the $\sim23$~arcmin lever arm between our new ACS fields and coordinated parallel observations with WFC3, we probe the outskirts of the Hercules dwarf and attempt to constrain a predicted distance gradient across the face of Hercules, expected if Hercules is elongated from tides \citep{MartinJin2010}. Fitting a linear gradient along the major-axis of Hercules, we find the best fit model to our data is $-0.10\pm0.48$~kpc/arcmin. This accuracy is insufficient to distinguish between competing models of Hercules, although the lack of a gradient is expected from \citet{Kupper2017}.
    \item Even with the deep {\it HST} imaging and {\it Gaia} proper motion information, our work shows that constraining a distance gradient in such a faint system is not trivial, and the thickness of the dwarf and field contamination introduce additional uncertainties. Therefore, we advocate for combined studies (e.g., deep photometry, multi-epoch spectroscopy, astrometric data), with the aid of dedicated theoretical work, to understand the true nature of Milky Way UFDs. 
\end{itemize}

Is Hercules a stellar stream in formation? It is hard to settle the question. Tidal destruction seems to be a viable option to explain its extreme ellipticity ($\sim0.7$) and observational evidence for association with a larger stream of stars \citep[e.g.,][]{Sand2009,Aden2009,Deason2012,Roderick2015,Garling2018}. On the other hand, \citet{Gregory2020} recently found no kinematic evidence in the form of a velocity gradient or velocity substructure. 
In this work, we address this problem with a complementary observational test, which is to search for a large distance gradient along its major-axis, as proposed by \citet{MartinJin2010}. Given the large uncertainties, our findings do not constrain the orbital models of Hercules. On the other hand, since the models proposed by  \citeauthor{Kupper2017} and \citeauthor{MartinJin2010} are both slightly incompatible with the observed proper motions of Hercules \citep{Gregory2020}, alternative explanations for its elongated shape such as formation through mergers or puffy dispersion-dominated disks \citep[e.g.,][]{Starkenburg2016,Wheeler2017} should also be explored. 
Considering that the overdensities identified in \citet{Sand2009} are likely associated with Hercules, deep wide-area photometric follow-up studies in the era of the Rubin Observatory and the Roman Space Telescope can serve as one of the most powerful ways to identify stellar streams and constrain tidal stripping in UFDs. In addition to in depth observational studies, theoretical studies looking at the tidal distortion of UFDs are needed to understand the nature of UFDs and ultimately put them into context with respect to the Cold Dark Matter paradigm for structure formation.

\acknowledgments
This research is based on observations made with the NASA/ESA Hubble Space Telescope obtained from the Space Telescope Science Institute, which is operated by the Association of Universities for Research in Astronomy, Inc., under NASA contract NAS 5–26555. These observations are associated with program 15182. EO was partially supported by NSF grant AST1815767. JS acknowledges support from a Packard Fellowship.

\vspace{5mm}
\facilities{HST (ACS, WFC3), Gaia}

\software{
The IDL Astronomy User's Library \citep{IDLforever},
DOLPHOT2.0 \citep{Dolphin2002},
Topcat \citep{Taylor2005}
}

\bibliographystyle{aasjournal}
\bibliography{reference}


\end{document}